\documentclass[10pt, aps, prl, superscriptaddress, amsmath, twocolumn, preprintnumbers]{revtex4-2}

\usepackage{xcolor}
\usepackage{dcolumn}
\usepackage{graphicx}
\usepackage{hyperref}

\begin{document}

\preprint{LLNL-JRNL-831022}

\title{Controlling extrapolations of nuclear properties with feature selection}


\author{Rodrigo Navarro P\'erez}
\email[]{rnavarroperez@sdsu.edu}
\affiliation{Department of Physics. San Diego State University. 5500 Campanile
 Drive, San Diego, California 02182-1233, USA}

\author{Nicolas Schunck}
\email[]{schunck1@llnl.gov}
\affiliation{Nuclear and Chemical Sciences Division, Lawrence Livermore
 National Laboratory, Livermore, California 94551, USA}


\date{\today}

\begin{abstract}

Predictions of nuclear properties far from measured data are inherently 
imprecise because of uncertainties in our knowledge of nuclear forces and in 
our treatment of quantum many-body effects in strongly-interacting systems.
While the model bias can be directly calculated when experimental 
data is available, only an estimate can be made in the absence of such 
measurements. Current approaches to compute the estimated bias quickly lose 
predictive power when input variables such as proton or neutron number are 
extrapolated, resulting in uncontrolled uncertainties in applications such as 
nucleosynthesis simulations. In this letter, we present a novel technique to 
identify the input variables of machine learning algorithms that can provide 
robust estimates of model bias. Our process is based on selecting input 
variables, or features, based on their probability distribution functions across the entire nuclear chart. 
We illustrate our approach on the problem of quantifying the model bias in 
nuclear binding energies calculated with Density Functional Theory (DFT). We 
show that feature selection can systematically improve theoretical predictions 
without increasing uncertainties.
\end{abstract}

\maketitle

{\em Introduction - }
Nuclear physics models are at the heart of some of the most important 
contemporary scientific problems. Prominent examples include probing the limits 
of the periodic table of elements 
\cite{giulianiColloquiumSuperheavyElements2019}, understanding the mechanisms 
of nuclear fission for basic science and carbon-free energy production sources 
\cite{schunckTheoryNuclearFission2022} or simulating the formation of 
elements in the universe through a set of nuclear reactions in various stellar 
environments \cite{cowanOriginHEaviestElements2021}. These problems require a 
complete description of nuclear properties for all nuclei across the nuclear 
chart, but only a very small subset of them have been measured experimentally. 
With the coming online of next-generation radioactive ion beam facilities, many 
of the gaps in our knowledge will be filled 
\cite{balantekinNuclearTheoryScience2014}. Yet, the vast majority of nuclear 
data will remain out of reach and will have to be determined through 
theoretical simulations.

Atomic nuclei are self-bound, strongly-interacting quantum many-body systems 
\cite{bohrNuclearStructure1998,ringNuclearManyBody2004}. The nuclear forces 
that bind them are not elementary interactions but are derived from 
non-perturbative quantum chromodynamics \cite{weinbergNuclearForcesChiral1990,
machleidtChiralEFTBased2016,hammerNuclearEffectiveField2020}. For these 
reasons, an exact description of the nucleus is impossible and approximations 
are needed \cite{hagenCoupledclusterComputationsAtomic2014,
carlsonQuantumMonteCarlo2015,hergertInMediumSimilarityRenormalization2016,
strobergNonempiricalInteractionsNuclear2019,schunckEnergyDensityFunctional2019}. 
As a result, all theories of the nucleus are in fact models characterized, 
among others, by a small number of parameters that must be calibrated to 
experimental data. Most importantly, the accuracy and precision of nuclear 
models tend to be far lower than what direct experimental measurements can 
deliver. For example, state-of-the-art mass models predict nuclear binding 
energies to within about 500 keV \cite{mollerNuclearGroundstateMasses2016,
gorielyHartreeFockBogoliubovNuclearMass2013} while Penning-trap 
mass measurements yield results within a few eV -- about 500 000 times more 
precise \cite{huangAME2020Atomic2021,wangAME2020Atomic2021}. In other words, 
nuclear models are inherently imperfect. Extrapolations far from the calibration 
region with such models are bound to come with a potentially very large bias.

The problems of calibrating nuclear model parameters, of quantifying and 
propagating the related uncertainties, and of estimating prediction biases have 
spurred interest in machine learning (ML) algorithms 
\cite{utamaNuclearMassPredictions2016, 
utamaRefiningMassFormulas2017,utamaValidatingNeuralnetworkRefinements2018,
neufcourtBayesianApproachModelbased2018, neufcourtNeutronDripLine2019,
kejzlarBayesianAveragingComputer2019, jiangExtrapolationNuclearStructure2019,
niuPredictionsNuclearBeta2019, wangBayesianEValuationIncomplete2019, 
lasseriTamingNuclearComplexity2020, keebleMachineLearningDeuteron2020}. 
These new applications have resulted in better estimates for the limits of the 
nuclear chart~\cite{neufcourtQuantifiedLimitsNuclear2020}, improved mass 
tables~\cite{utamaRefiningMassFormulas2017, 
utamaValidatingNeuralnetworkRefinements2018,
neufcourtBayesianApproachModelbased2018}, and extrapolations of no-core shell
model ground-state energies and radii to very large model spaces
\cite{jiangExtrapolationNuclearStructure2019}. A recent workshop 
on ``AI for nuclear physics'' presented an in-depth report of current and
possible applications of the broader field of artificial intelligence (AI) to
nuclear physics~\cite{bedaqueNuclearPhysics2021}. While AI/ML techniques 
have been used to correct imperfect nuclear models, the reliability of the 
resulting predictions far outside the range of measured data has been seldom 
addressed.

The usual approach in AI/ML starts by defining a specific relationship, 
represented as an algorithm, between a set of input variables (features) and a 
set of output variables (targets). This algorithm usually contains a set of 
adjustable parameters. Typical examples include 
Gaussian Processes, Support Vector Machines and Bayesian Neural Networks. The 
training of the algorithm is nothing but the determination of its parameters by 
minimizing the distance between the algorithm's output for a set of 
features-targets pairs and the physical world observations. In its most 
fundamental sense, this is similar to using an interpolating polynomial to
reproduce a set of given points.

Similar to polynomial interpolation, machine learning training can thus result 
in overfitting. This can be mitigated by randomly splitting the available 
physical world data into training and testing sets. The algorithm is trained 
with a subset of all available data and its performance is tested against the
remaining data. The training process is fine-tuned until the algorithm's
performance is comparable with the training and testing data. This process
ensures that the algorithm will make reliable predictions with new feature
values, as long as those values are within the training/testing domain
(i.e. interpolating). However there is no guarantee of reliable predictions
when the features are taken outside of the training/testing domain
(i.e. extrapolating). This would be akin to training an image recognition
software with pictures of dogs and wolfs, and then using it to identify the
animals in a picture of a lake with ducks and geese. For a simple and
pedagogical demonstration of the limits to extrapolations with an artificial
neural network, see~\cite{haleyExtrapolationLimitationsMultilayer1992}.

Nuclides throughout the nuclear chart are regularly identified by their number
of protons $Z$ and number of neutrons $N$. Practitioners of machine learning
in nuclear physics have, for the most part, adopted $Z$ and $N$ as the only
features. Making predictions for neutron-rich nuclei with AI/ML 
algorithms therefore implies that the value of these features is extrapolated 
outside the training/testing region. We will show below that such 
extrapolations are highly questionable. 

In this work we propose a practical approach to avoid extrapolations. The core 
of our solution consists in selecting features that stay within the 
training/testing domain when making predictions. Our procedure relies on three 
steps: (i) once a target has been established, define the training/testing 
region, (ii) define the prediction region, which will impact our choice of 
features for the training of the AI/ML algorithm. The choice of the prediction 
region depends on the problem at hand. For example, neutron-rich nuclei are 
essential for r-process simulations while a single isotopic sequence 
may be sufficient for nuclear structure studies, (iii) identify features with 
similar distributions in the training and prediction regions. This step will 
ensure that the algorithm is not extrapolated when making predictions.

{\em Theoretical Framework -}
Density Functional Theory (DFT) provides a particularly advantageous framework 
to implement our solution \cite{schunckEnergyDensityFunctional2019}. For every 
nuclide characterized by $Z$ and $N$, DFT calculations generate a large amount 
of properties, which range from physical observables like the binding energy or 
the neutron skin, to purely theoretical quantities such as the deformation 
parameters or the contribution to the binding energy from a specific term in an 
Energy Density Functional (EDF). Regardless of their nature, any of these 
properties can be used as a feature. We refer to the choice of $Z$ and $N$ only 
as the direct approach (DA). Including all available properties as features is 
called the all-properties approach (APA). Finally, to avoid extrapolations we 
propose to select as features those available DFT properties with similar 
distributions in the training and prediction regions. We refer to this option 
as the selected-properties approach (SPA) and will show that it is the only 
approach that shows consistently reliable performance.

In this work we evaluate how these three approaches improve the prediction of 
binding energies from two different parametrizations of the Skyrme energy 
density: SLY4 was optimized with an emphasis on neutron-rich nuclei and neutron 
matter properties~\cite{chabanatSkyrmeParametrizationSubnuclear1998} while UNEDF0 
focuses on properties of spherical and deformed nuclei and approximates particle 
number restoration with the Lipkin-Nogami(LN) prescription 
\cite{kortelainenNuclearEnergyDensity2010}. In the Supplemental material, we 
also include results obtained with the UNEDF1$_{\rm HFB}$ parametrization, 
which adds excitation energies of fission isomers to the UNEDF0 optimization 
protocol but drops the LN prescription and the center-of-mass correction 
\cite{schunckErrorAnalysisNuclear2015}.

Our target is the model bias in DFT binding energies, that is, the difference 
between the theoretical and real-world binding energy,
\begin{equation}
\Delta B(Z, N) = B_{\rm EDF}(Z, N) - B_{\rm phys}(Z,N).
\label{eq:bias}
\end{equation}
The bias is also referred to in the literature as the model discrepancy or
residual~\cite{jetscapecollaborationMultisystemBayesianConstraints2021,
utamaRefiningMassFormulas2017, neufcourtBayesianApproachModelbased2018}. The
goal is to obtain a reliable AI/ML estimate of the model bias
$\Delta B_{\rm ML}(\vec{f}(Z,N))$, where $\vec{f} = (f_1,\dots,f_{p})$ is a particular set of
$p$ features. Once the bias is estimated it can be subtracted from the bare DFT
calculations to obtain an improved set of binding energies
\begin{equation}
B^*(Z, N) = B_{\rm EDF}(Z,N) - \Delta B_{\rm ML}(\vec{f}(Z,N)).
\label{eq:removing_bias}
\end{equation}
To select the features with similar distributions in the training and
prediction regions we quantify the level of similarity between two
distributions from their overlap
\begin{equation}
BC(p,q) = \int \sqrt{p(x) q(x)} dx.
\label{eq:Bhattacharyya}
\end{equation}
This quantity is known as the Bhattacharyya coefficient~\cite
{bhattacharyyaMeasureDivergenceTwo1946} and has been used for feature
selection in classification problems~\cite
{niigakiCircularObjectDetection2012, patraNewSimilarityMeasure2015,
singhEnhancingRecommendationAccuracy2020,
vanmolleLeveragingBhattacharyyaCoefficient2021}. With this definition if $p
(x) = q(x)$ then $BC = 1$; if $p$ and $q$ do not overlap at all then $BC =
0$. For our purposes, the Bhattacharyya coefficient of each feature $f_i$
will be calculated with $p_i(x)$ being the probability distribution of $f_i$
in the training region and $q_i(x)$ the probability distribution in the
prediction region.

\begin{figure}[!htb]
\includegraphics[width=0.8\linewidth]{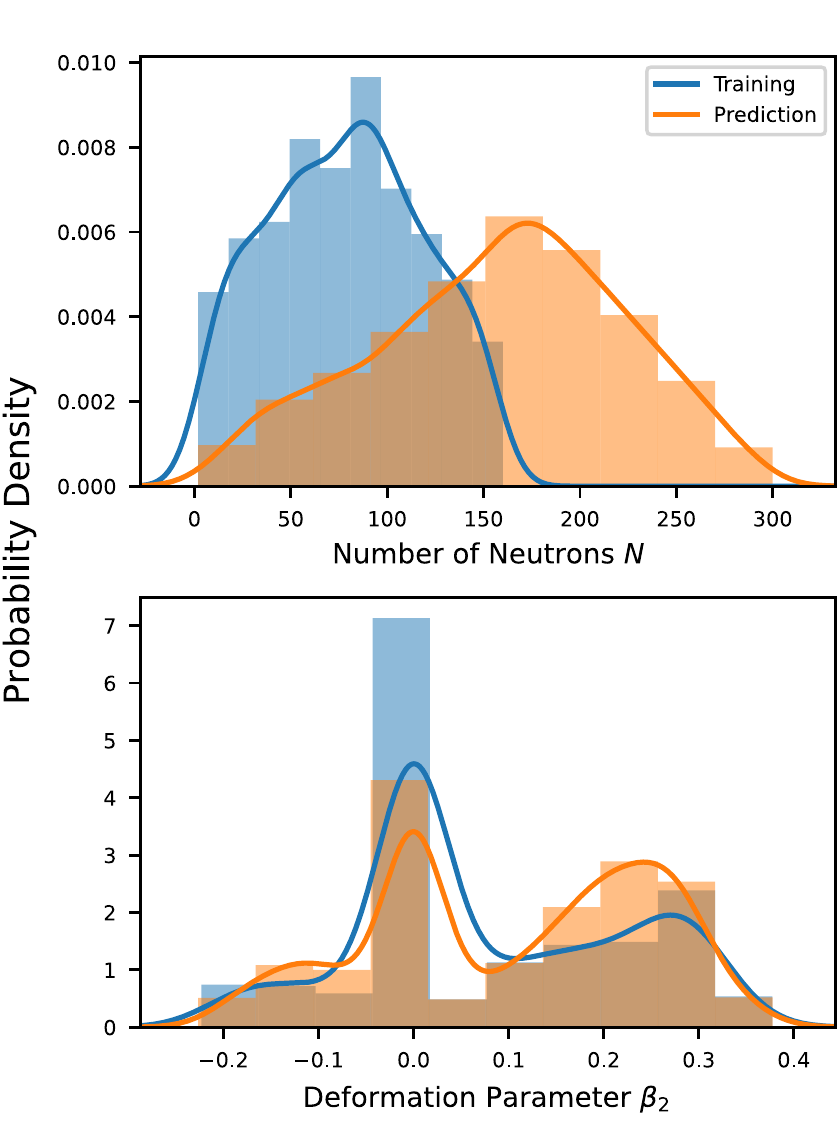}
\caption{\label{fig:overlaps} (Color on-line) Probability distribution
functions (solid lines) for the number of neutrons $N$ (top panel) and
the deformation parameter $\beta_2$ (bottom panel) calculated with
UNEDF0. The distributions are approximated using Kernel Density
Estimation from the available samples shown as histograms. Blue
corresponds to the data in the training region and orange corresponds to the
prediction region as described in the main text.}
\end{figure}

Mass tables of even-even nuclei were generated for each EDF using the
numerical solver HFBTHO \cite{navarroperezAxiallyDeformedSolution2017}. This 
means that we have \emph{samples} for each property instead of a probability 
distribution function. We employ Kernel Density Estimation (KDE)
\cite{rosenblattRemarksNonparametricEstimates1956,
parzenEstimationProbabilityDensity1962,gramackiNonparametricKernelDensity2017} 
to obtain a reliable representation of the probability distribution function of 
each property. Figure \ref{fig:overlaps} shows the samples and probability 
distribution functions for $N$ and for the deformation parameter $\beta_2$ 
calculated across the nuclear chart with the UNEDF0 functional. Blue 
corresponds to all even-even nuclei in the AME2020 mass evaluation data set
\cite{huangAME2020Atomic2021,wangAME2020Atomic2021}  and orange to 
all even-even nuclei between the proton and neutron driplines that are not in 
the AME2020 data set. The figure shows that using the neutron number $N$ 
($BC=0.56$) as a feature for training would require extrapolations when making 
predictions with the AI/ML algorithm. In contrast, using $\beta_2$ ($BC=0.94$) as 
a feature does not require an extrapolation since the distributions in the 
training and prediction regions are fairly similar. For the SPA, we use as 
features only those properties with $BC_i(p_{\rm train}, q_{\rm pred}) \ge c$, 
where $c$ is an arbitrary cut-off.

We use random forest regression to evaluate the performance of each feature 
selection method \cite{breimanRandomForests2001}. A random forest is a 
collection of decision trees where each tree is trained with a random subset of 
all training data. The output of the random forest is the average output of all 
decision trees. This technique avoids the overfitting that a few outliers can 
cause on a single decision tree. Additionally, random forests have few 
hyperparameters to tune, which makes them ideal for validation studies like the one 
presented here; for a complete introduction to random forest regression and
classification we refer the interested readers to 
\cite{biauRandomForestGuided2016} and \cite{genuerRandomForests2020}. Although 
the work presented here uses random forests, the concept of feature selection 
for extrapolation can be used with any other AI/ML algorithm.

\begin{figure}[t]
\includegraphics[width=\linewidth]{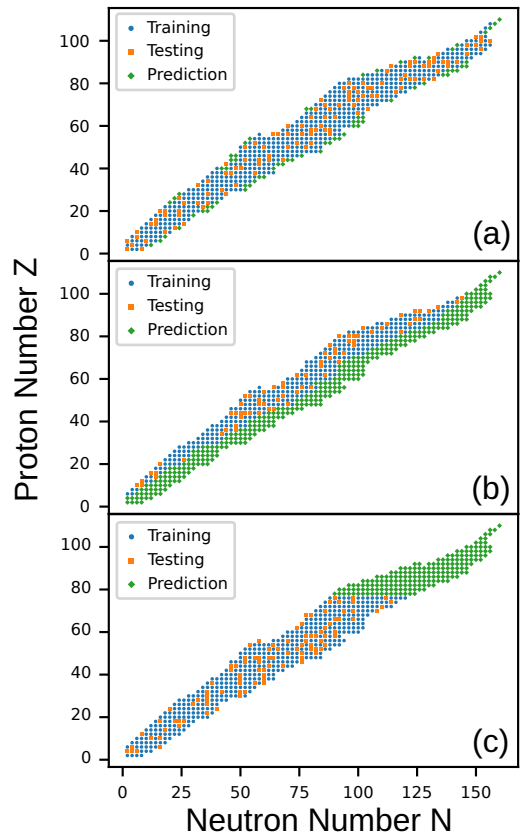}
\caption{\label{fig:3_landscapes} Nuclear landscapes indicating three
different splitting between training data (blue circles), testing data
(orange squares) and prediction data (green diamonds). Panel(a) shows
the time-wise split (TWS); panel (b) shows the large N split
(LNS); panel (c) shows the large Z split (LZS) as described in
the text.}
\end{figure}

\begin{figure*}[!ht]
\includegraphics[width=\linewidth]{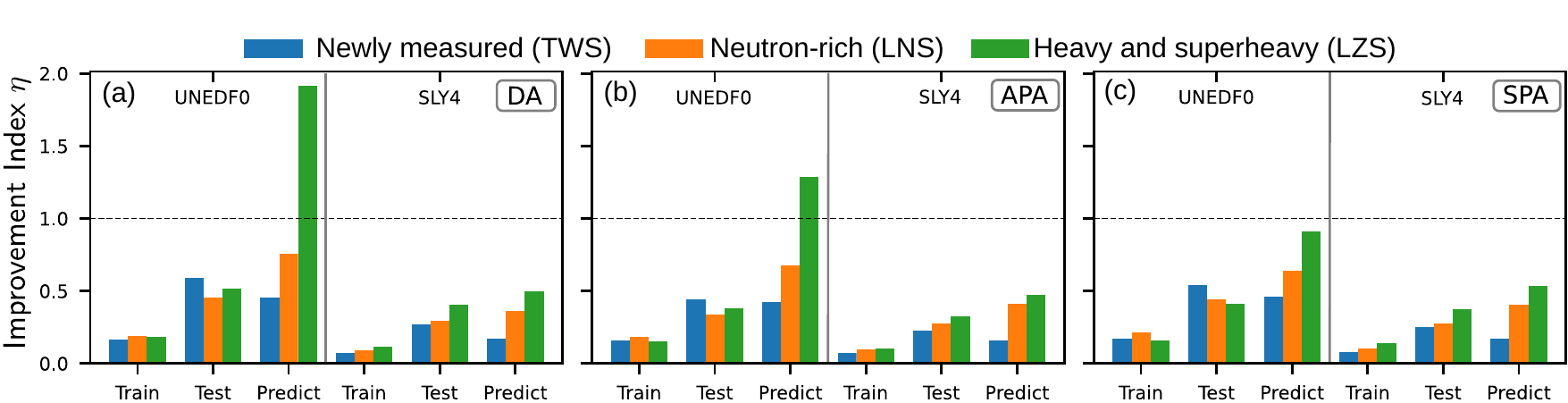}
\caption{\label{fig:rms_bars} Improvement index $\eta$ (see text) for 
the three feature selection approaches. Panel (a): direct approach (DA); 
panel (b): all-properties approach (APA); panel (c): selected-properties 
approach (SPA). Each panel shows results for the UNEDF0 and SLy4 
parametrizations of the Skyrme EDF and for the three regions: training, testing 
and prediction.}
\end{figure*}

{\em Results -}
We test the performance of our three feature selection approaches with 
available data. To this end, we define three prediction regions, which contain 
measured nuclei only, for which we will compare the estimated bias against the 
actual bias. The first region, dubbed time-wise split (TWS), consists of all the 
70 even-even nuclei in the AME2020 data set that are not in the AME2003 data 
set \cite{audiAME2003AtomicMass2003, wapstraAME2003AtomicMass2003}. This 
simulates a realistic scenario in which predictions are tested over time as new 
measurements become available but does not test predictions far away from the 
training data. The second region, dubbed large-N split (LNS), consists of the 
up to $6$ most neutron-rich even-even nuclei in each isotopic chain present in 
the AME2020 data set. This scenario tests predictions towards neutron-rich 
nuclei but still remains relatively close to the training data. The third 
region, called large-Z split (LZS), consists of all even-even nuclei in AME2020 
with $Z \ge 76$. This scenario will test predictions far away from the training 
data. In each case the training/testing region consists of all even-even nuclei 
in AME2020 that are not present in the corresponding prediction region. The 
split between training and testing data is performed by randomly selecting 
20\% nuclei for testing and leaving the remaining 80\% for training. Figure 
\ref{fig:3_landscapes} shows the training, testing and prediction regions as 
described above. Note that changing the prediction region changes the 
distribution of each property and it is thus necessary to recalculate the 
Bhattacharyya coefficient in order to re-select a new set of features with 
$BC_i \geq c$. As mentioned earlier the cut-off $c$ is somewhat arbitrary and 
an optimal value depends on the $BC$ value of all available properties. We set a
cut-off of $0.95$ and $0.90$ in the TWS and LNS cases respectively since most
properties have large $BC_i$ values: a lower cut-off would include most
properties as features and would not allow distinguishing between the APA and
SPA scenarios. For the LZS case we allow for a more ``lenient'' cut-off of
$0.75$ since there is a wider spread in the $BC_i$ values among all properties.

We quantify the performance of a theoretical model for binding energies by 
calculating the root mean square error (rmse) $\sigma = 
\sqrt{\frac{1}{n} \sum_{i=1}^n \left[B_{\rm theo}(Z_i, N_i) - B_{\rm exp}(Z_i, N_i) \right]^2}$,
where $n$ is the total number of nuclides in a particular region. To estimate 
the quality of the corrected model after removing the bias \eqref{eq:bias}, we 
then define the improvement index $\eta(\sigma) = \sigma / \sigma_0$ 
where $\sigma_{0}$ is the rmse of the original EDF model and $\sigma$ the one 
after removing the bias. Values $\eta < 1$ mean that removing the bias has 
reduced the rmse; values $\eta > 1$ mean that this modification has increased 
the rmse. Figure \ref{fig:rms_bars} shows the improvement index for the 
two EDFs considered in this work after removing the estimated bias with the 
three different feature selection methods. As expected, $\eta \ll 1$ in the 
training region, irrespective of the EDF, the type of data analyzed and the 
feature selection method. In the testing region, the improvement is less 
pronounced but in general $\eta$ stays lower than 0.5. Results in the 
prediction region are somewhat comparable to, or slightly worse than, in the testing region when
looking at data not too far from the training region (blue and orange) as already noticed in
\cite{neufcourtBayesianApproachModelbased2018}. When making predictions far 
from the training region (green), however, the improvement index takes values greater 
than 1 for the UNEDF0 functional for the DA and APA feature selection methods. 
This means that removing the AI/ML-estimated bias has in fact {\em degraded} the predictive power of 
the model. Similar results are obtained for the UNEDF1$_{\rm HFB}$ functional 
\cite{Supplemental}. Only by carefully selecting features with large $BC$ 
coefficients can we avoid this situation and keep $\eta < 1$. The case of SLy4 
sheds additional light on this mechanism. The error on binding energies comes 
mostly from not removing the effect of the electronic correction in the fit of 
the EDF. As a result, it is rather large and has a very clear dependency on 
$Z$; see Fig. 2 (b) in the Supplemental material \cite{Supplemental}. This 
trend can be easily learned by the AI/ML model, and the resulting bias estimate 
is robust (as a function of $Z$), accounts for most of the error and, 
therefore, still gives significant improvements in the prediction 
region, even far from training data. Put another way, feature selection is 
especially relevant for models with good predictive power with no clear the dependency of the model 
discrepancy on available properties.

{\em Conclusions -}
We have analyzed the performance of three feature selection approaches to
estimate the bias in DFT binding energies with two different EDFs. The 
performance of each approach was evaluated with three different splits of 
available data between training, testing and prediction. We showed that the 
direct approach, which uses $Z$ and $N$ as the only features, produces reliable 
predictions only for nuclei that are close to the available training data. When 
making predictions at values of $Z$ far away from the training data, removing 
the estimated bias can in fact reduce the accuracy of the DFT prediction. By 
selecting features with similar distributions in the training and prediction 
regions, our approach eliminates the need to extrapolate features when making 
predictions. It is the only one that consistently showed improvement among all 
three EDFs considered here and in all three splittings between training, 
testing and prediction data. While we demonstrated our feature-selection 
technique on nuclear binding energies computed from DFT, it could be 
generalized to any other observable and any other global theoretical model as
long as enough training data and potential features are available. In 
fact, it is also potentially applicable in every scientific discipline (i) that 
rely on approximate theoretical models with adjustable parameters and (ii) 
where extrapolations outside measured data are needed.

\begin{acknowledgments}
Discussions with M. Grosskopf, E. Lawrence, S. Wild and J. O'Neal are gratefully 
acknowledged.
Support for this work was partly provided through Scientific Discovery through
Advanced Computing (SciDAC) program funded by U.S. Department of Energy,
Office of Science, Advanced Scientific Computing Research and Nuclear
Physics. It was partly performed under the auspices of the US Department of
Energy by the Lawrence Livermore National Laboratory under Contract
DE-AC52-07NA27344. Computing support for this work came from the Lawrence
Livermore National Laboratory (LLNL) Institutional Computing Grand Challenge
program.
\end{acknowledgments}

\bibliography{book,zotero,references}

\newcommand{\noopsort}[1]{}
\begin{thebibliography}{51}%
\makeatletter
\providecommand \@ifxundefined [1]{%
 \@ifx{#1\undefined}
}%
\providecommand \@ifnum [1]{%
 \ifnum #1\expandafter \@firstoftwo
 \else \expandafter \@secondoftwo
 \fi
}%
\providecommand \@ifx [1]{%
 \ifx #1\expandafter \@firstoftwo
 \else \expandafter \@secondoftwo
 \fi
}%
\providecommand \natexlab [1]{#1}%
\providecommand \enquote  [1]{``#1''}%
\providecommand \bibnamefont  [1]{#1}%
\providecommand \bibfnamefont [1]{#1}%
\providecommand \citenamefont [1]{#1}%
\providecommand \href@noop [0]{\@secondoftwo}%
\providecommand \href [0]{\begingroup \@sanitize@url \@href}%
\providecommand \@href[1]{\@@startlink{#1}\@@href}%
\providecommand \@@href[1]{\endgroup#1\@@endlink}%
\providecommand \@sanitize@url [0]{\catcode `\\12\catcode `\$12\catcode
  `\&12\catcode `\#12\catcode `\^12\catcode `\_12\catcode `\%12\relax}%
\providecommand \@@startlink[1]{}%
\providecommand \@@endlink[0]{}%
\providecommand \url  [0]{\begingroup\@sanitize@url \@url }%
\providecommand \@url [1]{\endgroup\@href {#1}{\urlprefix }}%
\providecommand \urlprefix  [0]{URL }%
\providecommand \Eprint [0]{\href }%
\providecommand \doibase [0]{http://dx.doi.org/}%
\providecommand \selectlanguage [0]{\@gobble}%
\providecommand \bibinfo  [0]{\@secondoftwo}%
\providecommand \bibfield  [0]{\@secondoftwo}%
\providecommand \translation [1]{[#1]}%
\providecommand \BibitemOpen [0]{}%
\providecommand \bibitemStop [0]{}%
\providecommand \bibitemNoStop [0]{.\EOS\space}%
\providecommand \EOS [0]{\spacefactor3000\relax}%
\providecommand \BibitemShut  [1]{\csname bibitem#1\endcsname}%
\let\auto@bib@innerbib\@empty
\bibitem [{\citenamefont {Giuliani}\ \emph {et~al.}(2019)\citenamefont
  {Giuliani}, \citenamefont {Matheson}, \citenamefont {Nazarewicz},
  \citenamefont {Olsen}, \citenamefont {Reinhard}, \citenamefont {Sadhukhan},
  \citenamefont {Schuetrumpf}, \citenamefont {Schunck},\ and\ \citenamefont
  {Schwerdtfeger}}]{giulianiColloquiumSuperheavyElements2019}%
  \BibitemOpen
  \bibfield  {author} {\bibinfo {author} {\bibfnamefont {S.~A.}\ \bibnamefont
  {Giuliani}}, \bibinfo {author} {\bibfnamefont {Z.}~\bibnamefont {Matheson}},
  \bibinfo {author} {\bibfnamefont {W.}~\bibnamefont {Nazarewicz}}, \bibinfo
  {author} {\bibfnamefont {E.}~\bibnamefont {Olsen}}, \bibinfo {author}
  {\bibfnamefont {P.-G.}\ \bibnamefont {Reinhard}}, \bibinfo {author}
  {\bibfnamefont {J.}~\bibnamefont {Sadhukhan}}, \bibinfo {author}
  {\bibfnamefont {B.}~\bibnamefont {Schuetrumpf}}, \bibinfo {author}
  {\bibfnamefont {N.}~\bibnamefont {Schunck}}, \ and\ \bibinfo {author}
  {\bibfnamefont {P.}~\bibnamefont {Schwerdtfeger}},\ }\href {\doibase
  10.1103/RevModPhys.91.011001} {\bibfield  {journal} {\bibinfo  {journal}
  {Rev. Mod. Phys.}\ }\textbf {\bibinfo {volume} {91}},\ \bibinfo {pages}
  {011001} (\bibinfo {year} {2019})}\BibitemShut {NoStop}%
\bibitem [{\citenamefont {Schunck}\ and\ \citenamefont
  {Regnier}(2022)}]{schunckTheoryNuclearFission2022}%
  \BibitemOpen
  \bibfield  {author} {\bibinfo {author} {\bibfnamefont {N.}~\bibnamefont
  {Schunck}}\ and\ \bibinfo {author} {\bibfnamefont {D.}~\bibnamefont
  {Regnier}},\ }\href@noop {} {\enquote {\bibinfo {title} {Theory of nuclear
  fission},}\ } (\bibinfo {year} {2022}),\ \Eprint
  {http://arxiv.org/abs/2201.02719} {arXiv:2201.02719 [nucl-th]} \BibitemShut
  {NoStop}%
\bibitem [{\citenamefont {Cowan}\ \emph {et~al.}(2021)\citenamefont {Cowan},
  \citenamefont {Sneden}, \citenamefont {Lawler}, \citenamefont {Aprahamian},
  \citenamefont {Wiescher}, \citenamefont {Langanke}, \citenamefont
  {{Mart{\'i}nez-Pinedo}},\ and\ \citenamefont
  {Thielemann}}]{cowanOriginHEaviestElements2021}%
  \BibitemOpen
  \bibfield  {author} {\bibinfo {author} {\bibfnamefont {J.~J.}\ \bibnamefont
  {Cowan}}, \bibinfo {author} {\bibfnamefont {C.}~\bibnamefont {Sneden}},
  \bibinfo {author} {\bibfnamefont {J.~E.}\ \bibnamefont {Lawler}}, \bibinfo
  {author} {\bibfnamefont {A.}~\bibnamefont {Aprahamian}}, \bibinfo {author}
  {\bibfnamefont {M.}~\bibnamefont {Wiescher}}, \bibinfo {author}
  {\bibfnamefont {K.}~\bibnamefont {Langanke}}, \bibinfo {author}
  {\bibfnamefont {G.}~\bibnamefont {{Mart{\'i}nez-Pinedo}}}, \ and\ \bibinfo
  {author} {\bibfnamefont {F.-K.}\ \bibnamefont {Thielemann}},\ }\href
  {\doibase 10.1103/RevModPhys.93.015002} {\bibfield  {journal} {\bibinfo
  {journal} {Rev. Mod. Phys.}\ }\textbf {\bibinfo {volume} {93}},\ \bibinfo
  {pages} {015002} (\bibinfo {year} {2021})}\BibitemShut {NoStop}%
\bibitem [{\citenamefont {Balantekin}\ \emph {et~al.}(2014)\citenamefont
  {Balantekin}, \citenamefont {Carlson}, \citenamefont {Dean}, \citenamefont
  {Fuller}, \citenamefont {Furnstahl}, \citenamefont {{Hjorth-Jensen}},
  \citenamefont {Janssens}, \citenamefont {Li}, \citenamefont {Nazarewicz},
  \citenamefont {Nunes}, \citenamefont {Ormand}, \citenamefont {Reddy},\ and\
  \citenamefont {Sherrill}}]{balantekinNuclearTheoryScience2014}%
  \BibitemOpen
  \bibfield  {author} {\bibinfo {author} {\bibfnamefont {A.~B.}\ \bibnamefont
  {Balantekin}}, \bibinfo {author} {\bibfnamefont {J.}~\bibnamefont {Carlson}},
  \bibinfo {author} {\bibfnamefont {D.~J.}\ \bibnamefont {Dean}}, \bibinfo
  {author} {\bibfnamefont {G.~M.}\ \bibnamefont {Fuller}}, \bibinfo {author}
  {\bibfnamefont {R.~J.}\ \bibnamefont {Furnstahl}}, \bibinfo {author}
  {\bibfnamefont {M.}~\bibnamefont {{Hjorth-Jensen}}}, \bibinfo {author}
  {\bibfnamefont {R.~V.~F.}\ \bibnamefont {Janssens}}, \bibinfo {author}
  {\bibfnamefont {B.-A.}\ \bibnamefont {Li}}, \bibinfo {author} {\bibfnamefont
  {W.}~\bibnamefont {Nazarewicz}}, \bibinfo {author} {\bibfnamefont {F.~M.}\
  \bibnamefont {Nunes}}, \bibinfo {author} {\bibfnamefont {W.~E.}\ \bibnamefont
  {Ormand}}, \bibinfo {author} {\bibfnamefont {S.}~\bibnamefont {Reddy}}, \
  and\ \bibinfo {author} {\bibfnamefont {B.~M.}\ \bibnamefont {Sherrill}},\
  }\href {\doibase 10.1142/S0217732314300109} {\bibfield  {journal} {\bibinfo
  {journal} {Mod. Phys. Lett. A}\ }\textbf {\bibinfo {volume} {29}},\ \bibinfo
  {pages} {1430010} (\bibinfo {year} {2014})}\BibitemShut {NoStop}%
\bibitem [{\citenamefont {Bohr}\ and\ \citenamefont
  {Mottelson}(1998)}]{bohrNuclearStructure1998}%
  \BibitemOpen
  \bibfield  {author} {\bibinfo {author} {\bibfnamefont {A.}~\bibnamefont
  {Bohr}}\ and\ \bibinfo {author} {\bibfnamefont {B.}~\bibnamefont
  {Mottelson}},\ }\href {\doibase 10.1142/3530} {\emph {\bibinfo {title}
  {Nuclear Structure}}},\ Vol.\ \bibinfo {volume} {I, Single-Particle Motion}\
  (\bibinfo  {publisher} {World Scientific},\ \bibinfo {year}
  {1998})\BibitemShut {NoStop}%
\bibitem [{\citenamefont {Ring}\ and\ \citenamefont
  {Schuck}(2004)}]{ringNuclearManyBody2004}%
  \BibitemOpen
  \bibfield  {author} {\bibinfo {author} {\bibfnamefont {P.}~\bibnamefont
  {Ring}}\ and\ \bibinfo {author} {\bibfnamefont {P.}~\bibnamefont {Schuck}},\
  }\href@noop {} {\emph {\bibinfo {title} {The Nuclear Many-Body Problem}}},\
  Texts and Monographs in Physics\ (\bibinfo  {publisher} {Springer},\ \bibinfo
  {year} {2004})\BibitemShut {NoStop}%
\bibitem [{\citenamefont {Weinberg}(1990)}]{weinbergNuclearForcesChiral1990}%
  \BibitemOpen
  \bibfield  {author} {\bibinfo {author} {\bibfnamefont {S.}~\bibnamefont
  {Weinberg}},\ }\href {\doibase 10.1016/0370-2693(90)90938-3} {\bibfield
  {journal} {\bibinfo  {journal} {Phys. Lett. B}\ }\textbf {\bibinfo {volume}
  {251}},\ \bibinfo {pages} {288} (\bibinfo {year} {1990})}\BibitemShut
  {NoStop}%
\bibitem [{\citenamefont {Machleidt}\ and\ \citenamefont
  {Sammarruca}(2016)}]{machleidtChiralEFTBased2016}%
  \BibitemOpen
  \bibfield  {author} {\bibinfo {author} {\bibfnamefont {R.}~\bibnamefont
  {Machleidt}}\ and\ \bibinfo {author} {\bibfnamefont {F.}~\bibnamefont
  {Sammarruca}},\ }\href {\doibase 10.1088/0031-8949/91/8/083007} {\bibfield
  {journal} {\bibinfo  {journal} {Phys. Scr.}\ }\textbf {\bibinfo {volume}
  {91}},\ \bibinfo {pages} {083007} (\bibinfo {year} {2016})}\BibitemShut
  {NoStop}%
\bibitem [{\citenamefont {Hammer}\ \emph {et~al.}(2020)\citenamefont {Hammer},
  \citenamefont {K{\"o}nig},\ and\ \citenamefont {{\noopsort{kolck}}{van
  Kolck}}}]{hammerNuclearEffectiveField2020}%
  \BibitemOpen
  \bibfield  {author} {\bibinfo {author} {\bibfnamefont {H.-W.}\ \bibnamefont
  {Hammer}}, \bibinfo {author} {\bibfnamefont {S.}~\bibnamefont {K{\"o}nig}}, \
  and\ \bibinfo {author} {\bibfnamefont {U.}~\bibnamefont
  {{\noopsort{kolck}}{van Kolck}}},\ }\href {\doibase
  10.1103/RevModPhys.92.025004} {\bibfield  {journal} {\bibinfo  {journal}
  {Rev. Mod. Phys.}\ }\textbf {\bibinfo {volume} {92}},\ \bibinfo {pages}
  {025004} (\bibinfo {year} {2020})}\BibitemShut {NoStop}%
\bibitem [{\citenamefont {Hagen}\ \emph {et~al.}(2014)\citenamefont {Hagen},
  \citenamefont {Papenbrock}, \citenamefont {{Hjorth-Jensen}},\ and\
  \citenamefont {Dean}}]{hagenCoupledclusterComputationsAtomic2014}%
  \BibitemOpen
  \bibfield  {author} {\bibinfo {author} {\bibfnamefont {G.}~\bibnamefont
  {Hagen}}, \bibinfo {author} {\bibfnamefont {T.}~\bibnamefont {Papenbrock}},
  \bibinfo {author} {\bibfnamefont {M.}~\bibnamefont {{Hjorth-Jensen}}}, \ and\
  \bibinfo {author} {\bibfnamefont {D.~J.}\ \bibnamefont {Dean}},\ }\href
  {\doibase 10.1088/0034-4885/77/9/096302} {\bibfield  {journal} {\bibinfo
  {journal} {Rep. Prog. Phys.}\ }\textbf {\bibinfo {volume} {77}},\ \bibinfo
  {pages} {096302} (\bibinfo {year} {2014})}\BibitemShut {NoStop}%
\bibitem [{\citenamefont {Carlson}\ \emph {et~al.}(2015)\citenamefont
  {Carlson}, \citenamefont {Gandolfi}, \citenamefont {Pederiva}, \citenamefont
  {Pieper}, \citenamefont {Schiavilla}, \citenamefont {Schmidt},\ and\
  \citenamefont {Wiringa}}]{carlsonQuantumMonteCarlo2015}%
  \BibitemOpen
  \bibfield  {author} {\bibinfo {author} {\bibfnamefont {J.}~\bibnamefont
  {Carlson}}, \bibinfo {author} {\bibfnamefont {S.}~\bibnamefont {Gandolfi}},
  \bibinfo {author} {\bibfnamefont {F.}~\bibnamefont {Pederiva}}, \bibinfo
  {author} {\bibfnamefont {S.~C.}\ \bibnamefont {Pieper}}, \bibinfo {author}
  {\bibfnamefont {R.}~\bibnamefont {Schiavilla}}, \bibinfo {author}
  {\bibfnamefont {K.~E.}\ \bibnamefont {Schmidt}}, \ and\ \bibinfo {author}
  {\bibfnamefont {R.~B.}\ \bibnamefont {Wiringa}},\ }\href {\doibase
  10.1103/RevModPhys.87.1067} {\bibfield  {journal} {\bibinfo  {journal} {Rev.
  Mod. Phys.}\ }\textbf {\bibinfo {volume} {87}},\ \bibinfo {pages} {1067}
  (\bibinfo {year} {2015})}\BibitemShut {NoStop}%
\bibitem [{\citenamefont {Hergert}\ \emph {et~al.}(2016)\citenamefont
  {Hergert}, \citenamefont {Bogner}, \citenamefont {Morris}, \citenamefont
  {Schwenk},\ and\ \citenamefont
  {Tsukiyama}}]{hergertInMediumSimilarityRenormalization2016}%
  \BibitemOpen
  \bibfield  {author} {\bibinfo {author} {\bibfnamefont {H.}~\bibnamefont
  {Hergert}}, \bibinfo {author} {\bibfnamefont {S.~K.}\ \bibnamefont {Bogner}},
  \bibinfo {author} {\bibfnamefont {T.~D.}\ \bibnamefont {Morris}}, \bibinfo
  {author} {\bibfnamefont {A.}~\bibnamefont {Schwenk}}, \ and\ \bibinfo
  {author} {\bibfnamefont {K.}~\bibnamefont {Tsukiyama}},\ }\href {\doibase
  10.1016/j.physrep.2015.12.007} {\bibfield  {journal} {\bibinfo  {journal}
  {Phys. Rep.}\ }\bibinfo {series} {Memorial {{Volume}} in {{Honor}} of
  {{Gerald E}}. {{Brown}}},\ \textbf {\bibinfo {volume} {621}},\ \bibinfo
  {pages} {165} (\bibinfo {year} {2016})}\BibitemShut {NoStop}%
\bibitem [{\citenamefont {Stroberg}\ \emph {et~al.}(2019)\citenamefont
  {Stroberg}, \citenamefont {Hergert}, \citenamefont {Bogner},\ and\
  \citenamefont {Holt}}]{strobergNonempiricalInteractionsNuclear2019}%
  \BibitemOpen
  \bibfield  {author} {\bibinfo {author} {\bibfnamefont {S.~R.}\ \bibnamefont
  {Stroberg}}, \bibinfo {author} {\bibfnamefont {H.}~\bibnamefont {Hergert}},
  \bibinfo {author} {\bibfnamefont {S.~K.}\ \bibnamefont {Bogner}}, \ and\
  \bibinfo {author} {\bibfnamefont {J.~D.}\ \bibnamefont {Holt}},\ }\href
  {\doibase 10.1146/annurev-nucl-101917-021120} {\bibfield  {journal} {\bibinfo
   {journal} {Annu. Rev. Nucl. Part. Sci.}\ }\textbf {\bibinfo {volume} {69}},\
  \bibinfo {pages} {307} (\bibinfo {year} {2019})}\BibitemShut {NoStop}%
\bibitem [{\citenamefont {Schunck}\ \emph {et~al.}(2019)\citenamefont
  {Schunck}, \citenamefont {Bender}, \citenamefont {Bulgac}, \citenamefont
  {Duguet}, \citenamefont {Ebran}, \citenamefont {Engel}, \citenamefont
  {Michael}, \citenamefont {Kortelainen},\ and\ \citenamefont
  {Nakatsukasa}}]{schunckEnergyDensityFunctional2019}%
  \BibitemOpen
  \bibfield  {author} {\bibinfo {author} {\bibfnamefont {N.}~\bibnamefont
  {Schunck}}, \bibinfo {author} {\bibfnamefont {M.}~\bibnamefont {Bender}},
  \bibinfo {author} {\bibfnamefont {A.}~\bibnamefont {Bulgac}}, \bibinfo
  {author} {\bibfnamefont {T.}~\bibnamefont {Duguet}}, \bibinfo {author}
  {\bibfnamefont {J.-P.}\ \bibnamefont {Ebran}}, \bibinfo {author}
  {\bibfnamefont {J.}~\bibnamefont {Engel}}, \bibinfo {author} {\bibfnamefont
  {M.~F.}\ \bibnamefont {Michael}}, \bibinfo {author} {\bibfnamefont
  {M.}~\bibnamefont {Kortelainen}}, \ and\ \bibinfo {author} {\bibfnamefont
  {T.}~\bibnamefont {Nakatsukasa}},\ }\href@noop {} {\emph {\bibinfo {title}
  {Energy {{Density Functional Methods}} for {{Atomic Nuclei}}}}}\ (\bibinfo
  {publisher} {{Institute of Physics Publishing}},\ \bibinfo {year}
  {2019})\BibitemShut {NoStop}%
\bibitem [{\citenamefont {M{\"o}ller}\ \emph {et~al.}(2016)\citenamefont
  {M{\"o}ller}, \citenamefont {Sierk}, \citenamefont {Ichikawa},\ and\
  \citenamefont {Sagawa}}]{mollerNuclearGroundstateMasses2016}%
  \BibitemOpen
  \bibfield  {author} {\bibinfo {author} {\bibfnamefont {P.}~\bibnamefont
  {M{\"o}ller}}, \bibinfo {author} {\bibfnamefont {A.~J.}\ \bibnamefont
  {Sierk}}, \bibinfo {author} {\bibfnamefont {T.}~\bibnamefont {Ichikawa}}, \
  and\ \bibinfo {author} {\bibfnamefont {H.}~\bibnamefont {Sagawa}},\ }\href
  {\doibase 10.1016/j.adt.2015.10.002} {\bibfield  {journal} {\bibinfo
  {journal} {Atom. Data Nuc. Data Tab.}\ }\textbf {\bibinfo {volume} {109}},\
  \bibinfo {pages} {1} (\bibinfo {year} {2016})}\BibitemShut {NoStop}%
\bibitem [{\citenamefont {Goriely}\ \emph {et~al.}(2013)\citenamefont
  {Goriely}, \citenamefont {Chamel},\ and\ \citenamefont
  {Pearson}}]{gorielyHartreeFockBogoliubovNuclearMass2013}%
  \BibitemOpen
  \bibfield  {author} {\bibinfo {author} {\bibfnamefont {S.}~\bibnamefont
  {Goriely}}, \bibinfo {author} {\bibfnamefont {N.}~\bibnamefont {Chamel}}, \
  and\ \bibinfo {author} {\bibfnamefont {J.~M.}\ \bibnamefont {Pearson}},\
  }\href {\doibase 10.1103/PhysRevC.88.061302} {\bibfield  {journal} {\bibinfo
  {journal} {Phys. Rev. C}\ }\textbf {\bibinfo {volume} {88}},\ \bibinfo
  {pages} {061302} (\bibinfo {year} {2013})}\BibitemShut {NoStop}%
\bibitem [{\citenamefont {Huang}\ \emph {et~al.}(2021)\citenamefont {Huang},
  \citenamefont {Wang}, \citenamefont {Kondev}, \citenamefont {Audi},\ and\
  \citenamefont {Naimi}}]{huangAME2020Atomic2021}%
  \BibitemOpen
  \bibfield  {author} {\bibinfo {author} {\bibfnamefont {W.~J.}\ \bibnamefont
  {Huang}}, \bibinfo {author} {\bibfnamefont {M.}~\bibnamefont {Wang}},
  \bibinfo {author} {\bibfnamefont {F.~G.}\ \bibnamefont {Kondev}}, \bibinfo
  {author} {\bibfnamefont {G.}~\bibnamefont {Audi}}, \ and\ \bibinfo {author}
  {\bibfnamefont {S.}~\bibnamefont {Naimi}},\ }\href {\doibase
  10.1088/1674-1137/abddb0} {\bibfield  {journal} {\bibinfo  {journal} {Chinese
  Phys. C}\ }\textbf {\bibinfo {volume} {45}},\ \bibinfo {pages} {030002}
  (\bibinfo {year} {2021})}\BibitemShut {NoStop}%
\bibitem [{\citenamefont {Wang}\ \emph {et~al.}(2021)\citenamefont {Wang},
  \citenamefont {Huang}, \citenamefont {Kondev}, \citenamefont {Audi},\ and\
  \citenamefont {Naimi}}]{wangAME2020Atomic2021}%
  \BibitemOpen
  \bibfield  {author} {\bibinfo {author} {\bibfnamefont {M.}~\bibnamefont
  {Wang}}, \bibinfo {author} {\bibfnamefont {W.~J.}\ \bibnamefont {Huang}},
  \bibinfo {author} {\bibfnamefont {F.~G.}\ \bibnamefont {Kondev}}, \bibinfo
  {author} {\bibfnamefont {G.}~\bibnamefont {Audi}}, \ and\ \bibinfo {author}
  {\bibfnamefont {S.}~\bibnamefont {Naimi}},\ }\href {\doibase
  10.1088/1674-1137/abddaf} {\bibfield  {journal} {\bibinfo  {journal} {Chinese
  Phys. C}\ }\textbf {\bibinfo {volume} {45}},\ \bibinfo {pages} {030003}
  (\bibinfo {year} {2021})}\BibitemShut {NoStop}%
\bibitem [{\citenamefont {Utama}\ \emph {et~al.}(2016)\citenamefont {Utama},
  \citenamefont {Piekarewicz},\ and\ \citenamefont
  {Prosper}}]{utamaNuclearMassPredictions2016}%
  \BibitemOpen
  \bibfield  {author} {\bibinfo {author} {\bibfnamefont {R.}~\bibnamefont
  {Utama}}, \bibinfo {author} {\bibfnamefont {J.}~\bibnamefont {Piekarewicz}},
  \ and\ \bibinfo {author} {\bibfnamefont {H.~B.}\ \bibnamefont {Prosper}},\
  }\href {\doibase 10.1103/PhysRevC.93.014311} {\bibfield  {journal} {\bibinfo
  {journal} {Phys. Rev. C}\ }\textbf {\bibinfo {volume} {93}},\ \bibinfo
  {pages} {014311} (\bibinfo {year} {2016})}\BibitemShut {NoStop}%
\bibitem [{\citenamefont {Utama}\ and\ \citenamefont
  {Piekarewicz}(2017)}]{utamaRefiningMassFormulas2017}%
  \BibitemOpen
  \bibfield  {author} {\bibinfo {author} {\bibfnamefont {R.}~\bibnamefont
  {Utama}}\ and\ \bibinfo {author} {\bibfnamefont {J.}~\bibnamefont
  {Piekarewicz}},\ }\href {\doibase 10.1103/PhysRevC.96.044308} {\bibfield
  {journal} {\bibinfo  {journal} {Phys. Rev. C}\ }\textbf {\bibinfo {volume}
  {96}},\ \bibinfo {pages} {044308} (\bibinfo {year} {2017})}\BibitemShut
  {NoStop}%
\bibitem [{\citenamefont {Utama}\ and\ \citenamefont
  {Piekarewicz}(2018)}]{utamaValidatingNeuralnetworkRefinements2018}%
  \BibitemOpen
  \bibfield  {author} {\bibinfo {author} {\bibfnamefont {R.}~\bibnamefont
  {Utama}}\ and\ \bibinfo {author} {\bibfnamefont {J.}~\bibnamefont
  {Piekarewicz}},\ }\href {\doibase 10.1103/PhysRevC.97.014306} {\bibfield
  {journal} {\bibinfo  {journal} {Phys. Rev. C}\ }\textbf {\bibinfo {volume}
  {97}},\ \bibinfo {pages} {014306} (\bibinfo {year} {2018})}\BibitemShut
  {NoStop}%
\bibitem [{\citenamefont {Neufcourt}\ \emph {et~al.}(2018)\citenamefont
  {Neufcourt}, \citenamefont {Cao}, \citenamefont {Nazarewicz},\ and\
  \citenamefont {Viens}}]{neufcourtBayesianApproachModelbased2018}%
  \BibitemOpen
  \bibfield  {author} {\bibinfo {author} {\bibfnamefont {L.}~\bibnamefont
  {Neufcourt}}, \bibinfo {author} {\bibfnamefont {Y.}~\bibnamefont {Cao}},
  \bibinfo {author} {\bibfnamefont {W.}~\bibnamefont {Nazarewicz}}, \ and\
  \bibinfo {author} {\bibfnamefont {F.}~\bibnamefont {Viens}},\ }\href
  {\doibase 10.1103/PhysRevC.98.034318} {\bibfield  {journal} {\bibinfo
  {journal} {Phys. Rev. C}\ }\textbf {\bibinfo {volume} {98}},\ \bibinfo
  {pages} {034318} (\bibinfo {year} {2018})}\BibitemShut {NoStop}%
\bibitem [{\citenamefont {Neufcourt}\ \emph {et~al.}(2019)\citenamefont
  {Neufcourt}, \citenamefont {Cao}, \citenamefont {Nazarewicz}, \citenamefont
  {Olsen},\ and\ \citenamefont {Viens}}]{neufcourtNeutronDripLine2019}%
  \BibitemOpen
  \bibfield  {author} {\bibinfo {author} {\bibfnamefont {L.}~\bibnamefont
  {Neufcourt}}, \bibinfo {author} {\bibfnamefont {Y.}~\bibnamefont {Cao}},
  \bibinfo {author} {\bibfnamefont {W.}~\bibnamefont {Nazarewicz}}, \bibinfo
  {author} {\bibfnamefont {E.}~\bibnamefont {Olsen}}, \ and\ \bibinfo {author}
  {\bibfnamefont {F.}~\bibnamefont {Viens}},\ }\href {\doibase
  10.1103/PhysRevLett.122.062502} {\bibfield  {journal} {\bibinfo  {journal}
  {Phys. Rev. Lett.}\ }\textbf {\bibinfo {volume} {122}},\ \bibinfo {pages}
  {062502} (\bibinfo {year} {2019})}\BibitemShut {NoStop}%
\bibitem [{\citenamefont {Kejzlar}\ \emph {et~al.}(2019)\citenamefont
  {Kejzlar}, \citenamefont {Neufcourt}, \citenamefont {Maiti},\ and\
  \citenamefont {Viens}}]{kejzlarBayesianAveragingComputer2019}%
  \BibitemOpen
  \bibfield  {author} {\bibinfo {author} {\bibfnamefont {V.}~\bibnamefont
  {Kejzlar}}, \bibinfo {author} {\bibfnamefont {L.}~\bibnamefont {Neufcourt}},
  \bibinfo {author} {\bibfnamefont {T.}~\bibnamefont {Maiti}}, \ and\ \bibinfo
  {author} {\bibfnamefont {F.}~\bibnamefont {Viens}},\ }\href@noop {} {\
  (\bibinfo {year} {2019})},\ \Eprint {http://arxiv.org/abs/1904.04793}
  {arXiv:1904.04793} \BibitemShut {NoStop}%
\bibitem [{\citenamefont {Jiang}\ \emph {et~al.}(2019)\citenamefont {Jiang},
  \citenamefont {Hagen},\ and\ \citenamefont
  {Papenbrock}}]{jiangExtrapolationNuclearStructure2019}%
  \BibitemOpen
  \bibfield  {author} {\bibinfo {author} {\bibfnamefont {W.~G.}\ \bibnamefont
  {Jiang}}, \bibinfo {author} {\bibfnamefont {G.}~\bibnamefont {Hagen}}, \ and\
  \bibinfo {author} {\bibfnamefont {T.}~\bibnamefont {Papenbrock}},\ }\href
  {\doibase 10.1103/PhysRevC.100.054326} {\bibfield  {journal} {\bibinfo
  {journal} {Phys. Rev. C}\ }\textbf {\bibinfo {volume} {100}},\ \bibinfo
  {pages} {054326} (\bibinfo {year} {2019})}\BibitemShut {NoStop}%
\bibitem [{\citenamefont {Niu}\ \emph {et~al.}(2019)\citenamefont {Niu},
  \citenamefont {Liang}, \citenamefont {Sun}, \citenamefont {Long},\ and\
  \citenamefont {Niu}}]{niuPredictionsNuclearBeta2019}%
  \BibitemOpen
  \bibfield  {author} {\bibinfo {author} {\bibfnamefont {Z.~M.}\ \bibnamefont
  {Niu}}, \bibinfo {author} {\bibfnamefont {H.~Z.}\ \bibnamefont {Liang}},
  \bibinfo {author} {\bibfnamefont {B.~H.}\ \bibnamefont {Sun}}, \bibinfo
  {author} {\bibfnamefont {W.~H.}\ \bibnamefont {Long}}, \ and\ \bibinfo
  {author} {\bibfnamefont {Y.~F.}\ \bibnamefont {Niu}},\ }\href {\doibase
  10.1103/PhysRevC.99.064307} {\bibfield  {journal} {\bibinfo  {journal} {Phys.
  Rev. C}\ }\textbf {\bibinfo {volume} {99}},\ \bibinfo {pages} {064307}
  (\bibinfo {year} {2019})}\BibitemShut {NoStop}%
\bibitem [{\citenamefont {Wang}\ \emph {et~al.}(2019)\citenamefont {Wang},
  \citenamefont {Pei}, \citenamefont {Liu},\ and\ \citenamefont
  {Qiang}}]{wangBayesianEValuationIncomplete2019}%
  \BibitemOpen
  \bibfield  {author} {\bibinfo {author} {\bibfnamefont {Z.-A.}\ \bibnamefont
  {Wang}}, \bibinfo {author} {\bibfnamefont {J.}~\bibnamefont {Pei}}, \bibinfo
  {author} {\bibfnamefont {Y.}~\bibnamefont {Liu}}, \ and\ \bibinfo {author}
  {\bibfnamefont {Y.}~\bibnamefont {Qiang}},\ }\href {\doibase
  10.1103/PhysRevLett.123.122501} {\bibfield  {journal} {\bibinfo  {journal}
  {Phys. Rev. Lett.}\ }\textbf {\bibinfo {volume} {123}},\ \bibinfo {pages}
  {122501} (\bibinfo {year} {2019})}\BibitemShut {NoStop}%
\bibitem [{\citenamefont {Lasseri}\ \emph {et~al.}(2020)\citenamefont
  {Lasseri}, \citenamefont {Regnier}, \citenamefont {Ebran},\ and\
  \citenamefont {Penon}}]{lasseriTamingNuclearComplexity2020}%
  \BibitemOpen
  \bibfield  {author} {\bibinfo {author} {\bibfnamefont {R.-D.}\ \bibnamefont
  {Lasseri}}, \bibinfo {author} {\bibfnamefont {D.}~\bibnamefont {Regnier}},
  \bibinfo {author} {\bibfnamefont {J.-P.}\ \bibnamefont {Ebran}}, \ and\
  \bibinfo {author} {\bibfnamefont {A.}~\bibnamefont {Penon}},\ }\href
  {\doibase 10.1103/PhysRevLett.124.162502} {\bibfield  {journal} {\bibinfo
  {journal} {Phys. Rev. Lett.}\ }\textbf {\bibinfo {volume} {124}},\ \bibinfo
  {pages} {162502} (\bibinfo {year} {2020})}\BibitemShut {NoStop}%
\bibitem [{\citenamefont {Keeble}\ and\ \citenamefont
  {Rios}(2020)}]{keebleMachineLearningDeuteron2020}%
  \BibitemOpen
  \bibfield  {author} {\bibinfo {author} {\bibfnamefont {J.~W.~T.}\
  \bibnamefont {Keeble}}\ and\ \bibinfo {author} {\bibfnamefont
  {A.}~\bibnamefont {Rios}},\ }\href {\doibase 10.1016/j.physletb.2020.135743}
  {\bibfield  {journal} {\bibinfo  {journal} {Physics Letters B}\ }\textbf
  {\bibinfo {volume} {809}},\ \bibinfo {pages} {135743} (\bibinfo {year}
  {2020})}\BibitemShut {NoStop}%
\bibitem [{\citenamefont {Neufcourt}\ \emph {et~al.}(2020)\citenamefont
  {Neufcourt}, \citenamefont {Cao}, \citenamefont {Giuliani}, \citenamefont
  {Nazarewicz}, \citenamefont {Olsen},\ and\ \citenamefont
  {Tarasov}}]{neufcourtQuantifiedLimitsNuclear2020}%
  \BibitemOpen
  \bibfield  {author} {\bibinfo {author} {\bibfnamefont {L.}~\bibnamefont
  {Neufcourt}}, \bibinfo {author} {\bibfnamefont {Y.}~\bibnamefont {Cao}},
  \bibinfo {author} {\bibfnamefont {S.~A.}\ \bibnamefont {Giuliani}}, \bibinfo
  {author} {\bibfnamefont {W.}~\bibnamefont {Nazarewicz}}, \bibinfo {author}
  {\bibfnamefont {E.}~\bibnamefont {Olsen}}, \ and\ \bibinfo {author}
  {\bibfnamefont {O.~B.}\ \bibnamefont {Tarasov}},\ }\href {\doibase
  10.1103/PhysRevC.101.044307} {\bibfield  {journal} {\bibinfo  {journal}
  {Phys. Rev. C}\ }\textbf {\bibinfo {volume} {101}},\ \bibinfo {pages}
  {044307} (\bibinfo {year} {2020})}\BibitemShut {NoStop}%
\bibitem [{\citenamefont {Bedaque}\ \emph {et~al.}(2021)\citenamefont
  {Bedaque}, \citenamefont {Boehnlein}, \citenamefont {Cromaz}, \citenamefont
  {Diefenthaler}, \citenamefont {Elouadrhiri}, \citenamefont {Horn},
  \citenamefont {Kuchera}, \citenamefont {Lawrence}, \citenamefont {Lee},
  \citenamefont {Lidia}, \citenamefont {McKeown}, \citenamefont {Melnitchouk},
  \citenamefont {Nazarewicz}, \citenamefont {Orginos}, \citenamefont {Roblin},
  \citenamefont {Scott~Smith}, \citenamefont {Schram},\ and\ \citenamefont
  {Wang}}]{bedaqueNuclearPhysics2021}%
  \BibitemOpen
  \bibfield  {author} {\bibinfo {author} {\bibfnamefont {P.}~\bibnamefont
  {Bedaque}}, \bibinfo {author} {\bibfnamefont {A.}~\bibnamefont {Boehnlein}},
  \bibinfo {author} {\bibfnamefont {M.}~\bibnamefont {Cromaz}}, \bibinfo
  {author} {\bibfnamefont {M.}~\bibnamefont {Diefenthaler}}, \bibinfo {author}
  {\bibfnamefont {L.}~\bibnamefont {Elouadrhiri}}, \bibinfo {author}
  {\bibfnamefont {T.}~\bibnamefont {Horn}}, \bibinfo {author} {\bibfnamefont
  {M.}~\bibnamefont {Kuchera}}, \bibinfo {author} {\bibfnamefont
  {D.}~\bibnamefont {Lawrence}}, \bibinfo {author} {\bibfnamefont
  {D.}~\bibnamefont {Lee}}, \bibinfo {author} {\bibfnamefont {S.}~\bibnamefont
  {Lidia}}, \bibinfo {author} {\bibfnamefont {R.}~\bibnamefont {McKeown}},
  \bibinfo {author} {\bibfnamefont {W.}~\bibnamefont {Melnitchouk}}, \bibinfo
  {author} {\bibfnamefont {W.}~\bibnamefont {Nazarewicz}}, \bibinfo {author}
  {\bibfnamefont {K.}~\bibnamefont {Orginos}}, \bibinfo {author} {\bibfnamefont
  {Y.}~\bibnamefont {Roblin}}, \bibinfo {author} {\bibfnamefont
  {M.}~\bibnamefont {Scott~Smith}}, \bibinfo {author} {\bibfnamefont
  {M.}~\bibnamefont {Schram}}, \ and\ \bibinfo {author} {\bibfnamefont {X.-N.}\
  \bibnamefont {Wang}},\ }\href {\doibase 10.1140/epja/s10050-020-00290-x}
  {\bibfield  {journal} {\bibinfo  {journal} {Eur. Phys. J. A}\ }\textbf
  {\bibinfo {volume} {57}},\ \bibinfo {pages} {100} (\bibinfo {year}
  {2021})}\BibitemShut {NoStop}%
\bibitem [{\citenamefont {Haley}\ and\ \citenamefont
  {Soloway}(1992)}]{haleyExtrapolationLimitationsMultilayer1992}%
  \BibitemOpen
  \bibfield  {author} {\bibinfo {author} {\bibfnamefont {P.}~\bibnamefont
  {Haley}}\ and\ \bibinfo {author} {\bibfnamefont {D.}~\bibnamefont
  {Soloway}},\ }in\ \href {\doibase 10.1109/IJCNN.1992.227294} {\emph {\bibinfo
  {booktitle} {[{{Proceedings}} 1992] {{IJCNN International Joint Conference}}
  on {{Neural Networks}}}}},\ Vol.~\bibinfo {volume} {4}\ (\bibinfo {year}
  {1992})\ pp.\ \bibinfo {pages} {25--30 vol.4}\BibitemShut {NoStop}%
\bibitem [{\citenamefont {Chabanat}\ \emph {et~al.}(1998)\citenamefont
  {Chabanat}, \citenamefont {Bonche}, \citenamefont {Haensel}, \citenamefont
  {Meyer},\ and\ \citenamefont
  {Schaeffer}}]{chabanatSkyrmeParametrizationSubnuclear1998}%
  \BibitemOpen
  \bibfield  {author} {\bibinfo {author} {\bibfnamefont {E.}~\bibnamefont
  {Chabanat}}, \bibinfo {author} {\bibfnamefont {P.}~\bibnamefont {Bonche}},
  \bibinfo {author} {\bibfnamefont {P.}~\bibnamefont {Haensel}}, \bibinfo
  {author} {\bibfnamefont {J.}~\bibnamefont {Meyer}}, \ and\ \bibinfo {author}
  {\bibfnamefont {R.}~\bibnamefont {Schaeffer}},\ }\href {\doibase
  10.1016/S0375-9474(98)00180-8} {\bibfield  {journal} {\bibinfo  {journal}
  {Nucl. Phys. A}\ }\textbf {\bibinfo {volume} {635}},\ \bibinfo {pages} {231}
  (\bibinfo {year} {1998})}\BibitemShut {NoStop}%
\bibitem [{\citenamefont {Kortelainen}\ \emph {et~al.}(2010)\citenamefont
  {Kortelainen}, \citenamefont {Lesinski}, \citenamefont {Mor{\'e}},
  \citenamefont {Nazarewicz}, \citenamefont {Sarich}, \citenamefont {Schunck},
  \citenamefont {Stoitsov},\ and\ \citenamefont
  {Wild}}]{kortelainenNuclearEnergyDensity2010}%
  \BibitemOpen
  \bibfield  {author} {\bibinfo {author} {\bibfnamefont {M.}~\bibnamefont
  {Kortelainen}}, \bibinfo {author} {\bibfnamefont {T.}~\bibnamefont
  {Lesinski}}, \bibinfo {author} {\bibfnamefont {J.}~\bibnamefont {Mor{\'e}}},
  \bibinfo {author} {\bibfnamefont {W.}~\bibnamefont {Nazarewicz}}, \bibinfo
  {author} {\bibfnamefont {J.}~\bibnamefont {Sarich}}, \bibinfo {author}
  {\bibfnamefont {N.}~\bibnamefont {Schunck}}, \bibinfo {author} {\bibfnamefont
  {M.~V.}\ \bibnamefont {Stoitsov}}, \ and\ \bibinfo {author} {\bibfnamefont
  {S.}~\bibnamefont {Wild}},\ }\href {\doibase 10.1103/PhysRevC.82.024313}
  {\bibfield  {journal} {\bibinfo  {journal} {Phys. Rev. C}\ }\textbf {\bibinfo
  {volume} {82}},\ \bibinfo {pages} {024313} (\bibinfo {year}
  {2010})}\BibitemShut {NoStop}%
\bibitem [{\citenamefont {Schunck}\ \emph {et~al.}(2015)\citenamefont
  {Schunck}, \citenamefont {McDonnell}, \citenamefont {Sarich}, \citenamefont
  {Wild},\ and\ \citenamefont {Higdon}}]{schunckErrorAnalysisNuclear2015}%
  \BibitemOpen
  \bibfield  {author} {\bibinfo {author} {\bibfnamefont {N.}~\bibnamefont
  {Schunck}}, \bibinfo {author} {\bibfnamefont {J.~D.}\ \bibnamefont
  {McDonnell}}, \bibinfo {author} {\bibfnamefont {J.}~\bibnamefont {Sarich}},
  \bibinfo {author} {\bibfnamefont {S.~M.}\ \bibnamefont {Wild}}, \ and\
  \bibinfo {author} {\bibfnamefont {D.}~\bibnamefont {Higdon}},\ }\href
  {\doibase 10.1088/0954-3899/42/3/034024} {\bibfield  {journal} {\bibinfo
  {journal} {J. Phys. G: Nucl. Part. Phys.}\ }\textbf {\bibinfo {volume}
  {42}},\ \bibinfo {pages} {034024} (\bibinfo {year} {2015})}\BibitemShut
  {NoStop}%
\bibitem [{\citenamefont {{JETSCAPE Collaboration}}\ \emph
  {et~al.}(2021)\citenamefont {{JETSCAPE Collaboration}}, \citenamefont
  {Everett}, \citenamefont {Ke}, \citenamefont {Paquet}, \citenamefont
  {Vujanovic}, \citenamefont {Bass}, \citenamefont {Du}, \citenamefont {Gale},
  \citenamefont {Heffernan}, \citenamefont {Heinz}, \citenamefont {Liyanage},
  \citenamefont {Luzum}, \citenamefont {Majumder}, \citenamefont {McNelis},
  \citenamefont {Shen}, \citenamefont {Xu}, \citenamefont {Angerami},
  \citenamefont {Cao}, \citenamefont {Chen}, \citenamefont {Coleman},
  \citenamefont {Cunqueiro}, \citenamefont {Dai}, \citenamefont {Ehlers},
  \citenamefont {Elfner}, \citenamefont {Fan}, \citenamefont {Fries},
  \citenamefont {Garza}, \citenamefont {He}, \citenamefont {Jacak},
  \citenamefont {Jacobs}, \citenamefont {Jeon}, \citenamefont {Kim},
  \citenamefont {Kordell}, \citenamefont {Kumar}, \citenamefont {Mak},
  \citenamefont {Mulligan}, \citenamefont {Nattrass}, \citenamefont
  {Oliinychenko}, \citenamefont {Park}, \citenamefont {Putschke}, \citenamefont
  {Roland}, \citenamefont {Schenke}, \citenamefont {Schwiebert}, \citenamefont
  {Silva}, \citenamefont {Sirimanna}, \citenamefont {Soltz}, \citenamefont
  {Tachibana}, \citenamefont {Wang},\ and\ \citenamefont
  {Wolpert}}]{jetscapecollaborationMultisystemBayesianConstraints2021}%
  \BibitemOpen
  \bibfield  {author} {\bibinfo {author} {\bibnamefont {{JETSCAPE
  Collaboration}}}, \bibinfo {author} {\bibfnamefont {D.}~\bibnamefont
  {Everett}}, \bibinfo {author} {\bibfnamefont {W.}~\bibnamefont {Ke}},
  \bibinfo {author} {\bibfnamefont {J.-F.}\ \bibnamefont {Paquet}}, \bibinfo
  {author} {\bibfnamefont {G.}~\bibnamefont {Vujanovic}}, \bibinfo {author}
  {\bibfnamefont {S.~A.}\ \bibnamefont {Bass}}, \bibinfo {author}
  {\bibfnamefont {L.}~\bibnamefont {Du}}, \bibinfo {author} {\bibfnamefont
  {C.}~\bibnamefont {Gale}}, \bibinfo {author} {\bibfnamefont {M.}~\bibnamefont
  {Heffernan}}, \bibinfo {author} {\bibfnamefont {U.}~\bibnamefont {Heinz}},
  \bibinfo {author} {\bibfnamefont {D.}~\bibnamefont {Liyanage}}, \bibinfo
  {author} {\bibfnamefont {M.}~\bibnamefont {Luzum}}, \bibinfo {author}
  {\bibfnamefont {A.}~\bibnamefont {Majumder}}, \bibinfo {author}
  {\bibfnamefont {M.}~\bibnamefont {McNelis}}, \bibinfo {author} {\bibfnamefont
  {C.}~\bibnamefont {Shen}}, \bibinfo {author} {\bibfnamefont {Y.}~\bibnamefont
  {Xu}}, \bibinfo {author} {\bibfnamefont {A.}~\bibnamefont {Angerami}},
  \bibinfo {author} {\bibfnamefont {S.}~\bibnamefont {Cao}}, \bibinfo {author}
  {\bibfnamefont {Y.}~\bibnamefont {Chen}}, \bibinfo {author} {\bibfnamefont
  {J.}~\bibnamefont {Coleman}}, \bibinfo {author} {\bibfnamefont
  {L.}~\bibnamefont {Cunqueiro}}, \bibinfo {author} {\bibfnamefont
  {T.}~\bibnamefont {Dai}}, \bibinfo {author} {\bibfnamefont {R.}~\bibnamefont
  {Ehlers}}, \bibinfo {author} {\bibfnamefont {H.}~\bibnamefont {Elfner}},
  \bibinfo {author} {\bibfnamefont {W.}~\bibnamefont {Fan}}, \bibinfo {author}
  {\bibfnamefont {R.~J.}\ \bibnamefont {Fries}}, \bibinfo {author}
  {\bibfnamefont {F.}~\bibnamefont {Garza}}, \bibinfo {author} {\bibfnamefont
  {Y.}~\bibnamefont {He}}, \bibinfo {author} {\bibfnamefont {B.~V.}\
  \bibnamefont {Jacak}}, \bibinfo {author} {\bibfnamefont {P.~M.}\ \bibnamefont
  {Jacobs}}, \bibinfo {author} {\bibfnamefont {S.}~\bibnamefont {Jeon}},
  \bibinfo {author} {\bibfnamefont {B.}~\bibnamefont {Kim}}, \bibinfo {author}
  {\bibfnamefont {M.}~\bibnamefont {Kordell}}, \bibinfo {author} {\bibfnamefont
  {A.}~\bibnamefont {Kumar}}, \bibinfo {author} {\bibfnamefont
  {S.}~\bibnamefont {Mak}}, \bibinfo {author} {\bibfnamefont {J.}~\bibnamefont
  {Mulligan}}, \bibinfo {author} {\bibfnamefont {C.}~\bibnamefont {Nattrass}},
  \bibinfo {author} {\bibfnamefont {D.}~\bibnamefont {Oliinychenko}}, \bibinfo
  {author} {\bibfnamefont {C.}~\bibnamefont {Park}}, \bibinfo {author}
  {\bibfnamefont {J.~H.}\ \bibnamefont {Putschke}}, \bibinfo {author}
  {\bibfnamefont {G.}~\bibnamefont {Roland}}, \bibinfo {author} {\bibfnamefont
  {B.}~\bibnamefont {Schenke}}, \bibinfo {author} {\bibfnamefont
  {L.}~\bibnamefont {Schwiebert}}, \bibinfo {author} {\bibfnamefont
  {A.}~\bibnamefont {Silva}}, \bibinfo {author} {\bibfnamefont
  {C.}~\bibnamefont {Sirimanna}}, \bibinfo {author} {\bibfnamefont {R.~A.}\
  \bibnamefont {Soltz}}, \bibinfo {author} {\bibfnamefont {Y.}~\bibnamefont
  {Tachibana}}, \bibinfo {author} {\bibfnamefont {X.-N.}\ \bibnamefont {Wang}},
  \ and\ \bibinfo {author} {\bibfnamefont {R.~L.}\ \bibnamefont {Wolpert}},\
  }\href {\doibase 10.1103/PhysRevC.103.054904} {\bibfield  {journal} {\bibinfo
   {journal} {Phys. Rev. C}\ }\textbf {\bibinfo {volume} {103}},\ \bibinfo
  {pages} {054904} (\bibinfo {year} {2021})}\BibitemShut {NoStop}%
\bibitem [{\citenamefont
  {Bhattacharyya}(1946)}]{bhattacharyyaMeasureDivergenceTwo1946}%
  \BibitemOpen
  \bibfield  {author} {\bibinfo {author} {\bibfnamefont {A.}~\bibnamefont
  {Bhattacharyya}},\ }\href@noop {} {\bibfield  {journal} {\bibinfo  {journal}
  {Sankhy\=a: The Indian Journal of Statistics (1933-1960)}\ }\textbf {\bibinfo
  {volume} {7}},\ \bibinfo {pages} {401} (\bibinfo {year} {1946})}\BibitemShut
  {NoStop}%
\bibitem [{\citenamefont {Niigaki}\ \emph {et~al.}(2012)\citenamefont
  {Niigaki}, \citenamefont {Shimamura},\ and\ \citenamefont
  {Morimoto}}]{niigakiCircularObjectDetection2012}%
  \BibitemOpen
  \bibfield  {author} {\bibinfo {author} {\bibfnamefont {H.}~\bibnamefont
  {Niigaki}}, \bibinfo {author} {\bibfnamefont {J.}~\bibnamefont {Shimamura}},
  \ and\ \bibinfo {author} {\bibfnamefont {M.}~\bibnamefont {Morimoto}},\ }in\
  \href@noop {} {\emph {\bibinfo {booktitle} {Proceedings of the 21st
  {{International Conference}} on {{Pattern Recognition}} ({{ICPR2012}})}}}\
  (\bibinfo {year} {2012})\ pp.\ \bibinfo {pages} {2009--2012}\BibitemShut
  {NoStop}%
\bibitem [{\citenamefont {Patra}\ \emph {et~al.}(2015)\citenamefont {Patra},
  \citenamefont {Launonen}, \citenamefont {Ollikainen},\ and\ \citenamefont
  {Nandi}}]{patraNewSimilarityMeasure2015}%
  \BibitemOpen
  \bibfield  {author} {\bibinfo {author} {\bibfnamefont {B.~K.}\ \bibnamefont
  {Patra}}, \bibinfo {author} {\bibfnamefont {R.}~\bibnamefont {Launonen}},
  \bibinfo {author} {\bibfnamefont {V.}~\bibnamefont {Ollikainen}}, \ and\
  \bibinfo {author} {\bibfnamefont {S.}~\bibnamefont {Nandi}},\ }\href
  {\doibase 10.1016/j.knosys.2015.03.001} {\bibfield  {journal} {\bibinfo
  {journal} {Knowledge-Based Systems}\ }\textbf {\bibinfo {volume} {82}},\
  \bibinfo {pages} {163} (\bibinfo {year} {2015})}\BibitemShut {NoStop}%
\bibitem [{\citenamefont {Singh}\ \emph {et~al.}(2020)\citenamefont {Singh},
  \citenamefont {Sinha}, \citenamefont {Das},\ and\ \citenamefont
  {Choudhury}}]{singhEnhancingRecommendationAccuracy2020}%
  \BibitemOpen
  \bibfield  {author} {\bibinfo {author} {\bibfnamefont {P.~K.}\ \bibnamefont
  {Singh}}, \bibinfo {author} {\bibfnamefont {M.}~\bibnamefont {Sinha}},
  \bibinfo {author} {\bibfnamefont {S.}~\bibnamefont {Das}}, \ and\ \bibinfo
  {author} {\bibfnamefont {P.}~\bibnamefont {Choudhury}},\ }\href {\doibase
  10.1007/s10489-020-01775-4} {\bibfield  {journal} {\bibinfo  {journal} {Appl
  Intell}\ }\textbf {\bibinfo {volume} {50}},\ \bibinfo {pages} {4708}
  (\bibinfo {year} {2020})}\BibitemShut {NoStop}%
\bibitem [{\citenamefont {Van~Molle}\ \emph {et~al.}(2021)\citenamefont
  {Van~Molle}, \citenamefont {Verbelen}, \citenamefont {Vankeirsbilck},
  \citenamefont {De~Vylder}, \citenamefont {Diricx}, \citenamefont {Kimpe},
  \citenamefont {Simoens},\ and\ \citenamefont
  {Dhoedt}}]{vanmolleLeveragingBhattacharyyaCoefficient2021}%
  \BibitemOpen
  \bibfield  {author} {\bibinfo {author} {\bibfnamefont {P.}~\bibnamefont
  {Van~Molle}}, \bibinfo {author} {\bibfnamefont {T.}~\bibnamefont {Verbelen}},
  \bibinfo {author} {\bibfnamefont {B.}~\bibnamefont {Vankeirsbilck}}, \bibinfo
  {author} {\bibfnamefont {J.}~\bibnamefont {De~Vylder}}, \bibinfo {author}
  {\bibfnamefont {B.}~\bibnamefont {Diricx}}, \bibinfo {author} {\bibfnamefont
  {T.}~\bibnamefont {Kimpe}}, \bibinfo {author} {\bibfnamefont
  {P.}~\bibnamefont {Simoens}}, \ and\ \bibinfo {author} {\bibfnamefont
  {B.}~\bibnamefont {Dhoedt}},\ }\href {\doibase 10.1007/s00521-021-05789-y}
  {\bibfield  {journal} {\bibinfo  {journal} {Neural Comput \& Applic}\
  }\textbf {\bibinfo {volume} {33}},\ \bibinfo {pages} {10259} (\bibinfo {year}
  {2021})}\BibitemShut {NoStop}%
\bibitem [{\citenamefont {Navarro~P{\'e}rez}\ \emph {et~al.}(2017)\citenamefont
  {Navarro~P{\'e}rez}, \citenamefont {Schunck}, \citenamefont {Lasseri},
  \citenamefont {Zhang},\ and\ \citenamefont
  {Sarich}}]{navarroperezAxiallyDeformedSolution2017}%
  \BibitemOpen
  \bibfield  {author} {\bibinfo {author} {\bibfnamefont {R.}~\bibnamefont
  {Navarro~P{\'e}rez}}, \bibinfo {author} {\bibfnamefont {N.}~\bibnamefont
  {Schunck}}, \bibinfo {author} {\bibfnamefont {R.~D.}\ \bibnamefont
  {Lasseri}}, \bibinfo {author} {\bibfnamefont {C.}~\bibnamefont {Zhang}}, \
  and\ \bibinfo {author} {\bibfnamefont {J.}~\bibnamefont {Sarich}},\ }\href
  {\doibase 10.1016/j.cpc.2017.06.022} {\bibfield  {journal} {\bibinfo
  {journal} {Computer Physics Communications}\ }\textbf {\bibinfo {volume}
  {220}},\ \bibinfo {pages} {363} (\bibinfo {year} {2017})}\BibitemShut
  {NoStop}%
\bibitem [{\citenamefont
  {Rosenblatt}(1956)}]{rosenblattRemarksNonparametricEstimates1956}%
  \BibitemOpen
  \bibfield  {author} {\bibinfo {author} {\bibfnamefont {M.}~\bibnamefont
  {Rosenblatt}},\ }\href {\doibase 10.1214/aoms/1177728190} {\bibfield
  {journal} {\bibinfo  {journal} {The Annals of Mathematical Statistics}\
  }\textbf {\bibinfo {volume} {27}},\ \bibinfo {pages} {832} (\bibinfo {year}
  {1956})}\BibitemShut {NoStop}%
\bibitem [{\citenamefont
  {Parzen}(1962)}]{parzenEstimationProbabilityDensity1962}%
  \BibitemOpen
  \bibfield  {author} {\bibinfo {author} {\bibfnamefont {E.}~\bibnamefont
  {Parzen}},\ }\href {\doibase 10.1214/aoms/1177704472} {\bibfield  {journal}
  {\bibinfo  {journal} {The Annals of Mathematical Statistics}\ }\textbf
  {\bibinfo {volume} {33}},\ \bibinfo {pages} {1065} (\bibinfo {year}
  {1962})}\BibitemShut {NoStop}%
\bibitem [{\citenamefont
  {Gramacki}(2017)}]{gramackiNonparametricKernelDensity2017}%
  \BibitemOpen
  \bibfield  {author} {\bibinfo {author} {\bibfnamefont {A.}~\bibnamefont
  {Gramacki}},\ }\href@noop {} {\emph {\bibinfo {title} {Nonparametric {{Kernel
  Density Estimation}} and {{Its Computational Aspects}}}}}\ (\bibinfo
  {publisher} {{Springer}},\ \bibinfo {year} {2017})\BibitemShut {NoStop}%
\bibitem [{\citenamefont {Breiman}(2001)}]{breimanRandomForests2001}%
  \BibitemOpen
  \bibfield  {author} {\bibinfo {author} {\bibfnamefont {L.}~\bibnamefont
  {Breiman}},\ }\href {\doibase 10.1023/A:1010933404324} {\bibfield  {journal}
  {\bibinfo  {journal} {Machine Learning}\ }\textbf {\bibinfo {volume} {45}},\
  \bibinfo {pages} {5} (\bibinfo {year} {2001})}\BibitemShut {NoStop}%
\bibitem [{\citenamefont {Biau}\ and\ \citenamefont
  {Scornet}(2016)}]{biauRandomForestGuided2016}%
  \BibitemOpen
  \bibfield  {author} {\bibinfo {author} {\bibfnamefont {G.}~\bibnamefont
  {Biau}}\ and\ \bibinfo {author} {\bibfnamefont {E.}~\bibnamefont {Scornet}},\
  }\href {\doibase 10.1007/s11749-016-0481-7} {\bibfield  {journal} {\bibinfo
  {journal} {TEST}\ }\textbf {\bibinfo {volume} {25}},\ \bibinfo {pages} {197}
  (\bibinfo {year} {2016})}\BibitemShut {NoStop}%
\bibitem [{\citenamefont {Genuer}\ and\ \citenamefont
  {Poggi}(2020)}]{genuerRandomForests2020}%
  \BibitemOpen
  \bibfield  {author} {\bibinfo {author} {\bibfnamefont {R.}~\bibnamefont
  {Genuer}}\ and\ \bibinfo {author} {\bibfnamefont {J.-M.}\ \bibnamefont
  {Poggi}},\ }\href@noop {} {\emph {\bibinfo {title} {Random {{Forests}} with
  {{R}}}}}\ (\bibinfo  {publisher} {{Springer Nature}},\ \bibinfo {year}
  {2020})\BibitemShut {NoStop}%
\bibitem [{\citenamefont {Audi}\ \emph {et~al.}(2003)\citenamefont {Audi},
  \citenamefont {Wapstra},\ and\ \citenamefont
  {Thibault}}]{audiAME2003AtomicMass2003}%
  \BibitemOpen
  \bibfield  {author} {\bibinfo {author} {\bibfnamefont {G.}~\bibnamefont
  {Audi}}, \bibinfo {author} {\bibfnamefont {A.~H.}\ \bibnamefont {Wapstra}}, \
  and\ \bibinfo {author} {\bibfnamefont {C.}~\bibnamefont {Thibault}},\ }\href
  {\doibase 10.1016/j.nuclphysa.2003.11.003} {\bibfield  {journal} {\bibinfo
  {journal} {Nucl. Phys. A}\ }\bibinfo {series} {The 2003 {{NUBASE}} and
  {{Atomic Mass Evaluations}}},\ \textbf {\bibinfo {volume} {729}},\ \bibinfo
  {pages} {337} (\bibinfo {year} {2003})}\BibitemShut {NoStop}%
\bibitem [{\citenamefont {Wapstra}\ \emph {et~al.}(2003)\citenamefont
  {Wapstra}, \citenamefont {Audi},\ and\ \citenamefont
  {Thibault}}]{wapstraAME2003AtomicMass2003}%
  \BibitemOpen
  \bibfield  {author} {\bibinfo {author} {\bibfnamefont {A.~H.}\ \bibnamefont
  {Wapstra}}, \bibinfo {author} {\bibfnamefont {G.}~\bibnamefont {Audi}}, \
  and\ \bibinfo {author} {\bibfnamefont {C.}~\bibnamefont {Thibault}},\ }\href
  {\doibase 10.1016/j.nuclphysa.2003.11.002} {\bibfield  {journal} {\bibinfo
  {journal} {Nucl. Sci. Eng.}\ }\bibinfo {series} {The 2003 {{NUBASE}} and
  {{Atomic Mass Evaluations}}},\ \textbf {\bibinfo {volume} {729}},\ \bibinfo
  {pages} {129} (\bibinfo {year} {2003})}\BibitemShut {NoStop}%
\bibitem [{Sup()}]{Supplemental}%
  \BibitemOpen
  \href@noop {} {}\bibinfo {note} {See Supplemental Material}\BibitemShut
  {NoStop}%
\end{thebibliography}%


\begin{thebibliography}{1}%
\makeatletter
\providecommand \@ifxundefined [1]{%
 \@ifx{#1\undefined}
}%
\providecommand \@ifnum [1]{%
 \ifnum #1\expandafter \@firstoftwo
 \else \expandafter \@secondoftwo
 \fi
}%
\providecommand \@ifx [1]{%
 \ifx #1\expandafter \@firstoftwo
 \else \expandafter \@secondoftwo
 \fi
}%
\providecommand \natexlab [1]{#1}%
\providecommand \enquote  [1]{``#1''}%
\providecommand \bibnamefont  [1]{#1}%
\providecommand \bibfnamefont [1]{#1}%
\providecommand \citenamefont [1]{#1}%
\providecommand \href@noop [0]{\@secondoftwo}%
\providecommand \href [0]{\begingroup \@sanitize@url \@href}%
\providecommand \@href[1]{\@@startlink{#1}\@@href}%
\providecommand \@@href[1]{\endgroup#1\@@endlink}%
\providecommand \@sanitize@url [0]{\catcode `\\12\catcode `\$12\catcode
  `\&12\catcode `\#12\catcode `\^12\catcode `\_12\catcode `\%12\relax}%
\providecommand \@@startlink[1]{}%
\providecommand \@@endlink[0]{}%
\providecommand \url  [0]{\begingroup\@sanitize@url \@url }%
\providecommand \@url [1]{\endgroup\@href {#1}{\urlprefix }}%
\providecommand \urlprefix  [0]{URL }%
\providecommand \Eprint [0]{\href }%
\providecommand \doibase [0]{http://dx.doi.org/}%
\providecommand \selectlanguage [0]{\@gobble}%
\providecommand \bibinfo  [0]{\@secondoftwo}%
\providecommand \bibfield  [0]{\@secondoftwo}%
\providecommand \translation [1]{[#1]}%
\providecommand \BibitemOpen [0]{}%
\providecommand \bibitemStop [0]{}%
\providecommand \bibitemNoStop [0]{.\EOS\space}%
\providecommand \EOS [0]{\spacefactor3000\relax}%
\providecommand \BibitemShut  [1]{\csname bibitem#1\endcsname}%
\let\auto@bib@innerbib\@empty
\bibitem [{\citenamefont {Navarro~P{\'e}rez}\ \emph {et~al.}(2017)\citenamefont
  {Navarro~P{\'e}rez}, \citenamefont {Schunck}, \citenamefont {Lasseri},
  \citenamefont {Zhang},\ and\ \citenamefont
  {Sarich}}]{navarroperezAxiallyDeformedSolution2017}%
  \BibitemOpen
  \bibfield  {author} {\bibinfo {author} {\bibfnamefont {R.}~\bibnamefont
  {Navarro~P{\'e}rez}}, \bibinfo {author} {\bibfnamefont {N.}~\bibnamefont
  {Schunck}}, \bibinfo {author} {\bibfnamefont {R.~D.}\ \bibnamefont
  {Lasseri}}, \bibinfo {author} {\bibfnamefont {C.}~\bibnamefont {Zhang}}, \
  and\ \bibinfo {author} {\bibfnamefont {J.}~\bibnamefont {Sarich}},\ }\href
  {\doibase 10.1016/j.cpc.2017.06.022} {\bibfield  {journal} {\bibinfo
  {journal} {Computer Physics Communications}\ }\textbf {\bibinfo {volume}
  {220}},\ \bibinfo {pages} {363} (\bibinfo {year} {2017})}\BibitemShut
  {NoStop}%
\end{thebibliography}%

\end{document}


\preprint{LLNL-JRNL-831022}

\title{Supplemental Material to ``Controlling extrapolations of nuclear properties with feature selection''}


\author{Rodrigo Navarro P\'erez}
\email[]{rnavarroperez@sdsu.edu}
\affiliation{Department of Physics. San Diego State University. 5500 Campanile
 Drive, San Diego, California 02182-1233, USA}

\author{Nicolas Schunck}
\email[]{schunck1@llnl.gov}
\affiliation{Nuclear and Chemical Sciences Division, Lawrence Livermore
 National Laboratory, Livermore, California 94551, USA}


\date{\today}

\maketitle


\section{Calculations of HFBTHO Mass Tables}

Calculations of nuclear binding energies for even-even nuclei were performed 
with the code HFBTHO \cite{navarroperezAxiallyDeformedSolution2017} using an 
axially-deformed basis with $N_{\rm 0} = 20$ shells. The oscillator length was 
$b_0 = \sqrt{\hbar^2 c^2/(mc^2 \omega_0)}$ fm, where $mc^2 = 931.494013$ MeV, 
$\omega_0 = 1.2 \times 41 A^{-1/3}$ and $A$ is the mass number of the nucleus. 
Pairing correlations were treated at the HFB approximation (SLy4, 
UNEDF1$_{\rm HFB}$) and HFB with Lipkin-Nogami (UNEDF0). In all cases, the 
pairing interaction was a mixed surface-volume, density-dependent pairing 
force. In the calculation of the densities, only quasiparticles with energies 
less than $E_{\rm cut} = 60$ MeV were included. In the case of the SLy4 
interaction, the pairing strengths were set to $V_0^{(n)} = -325.25$ MeV and 
$V_0^{(p)} = -340.0625$ MeV. In every nucleus, the ground state was identified 
by performing calculations constrained on the axial quadrupole moment $Q_2$ for 
a few iterations before the constraint was released to identify the nearest 
unconstrained minimum. The value of this temporary constraint was 
$Q_2 = \beta_2 \sqrt{\frac{5}{\pi}}A^{5/3}$ where $\beta_2$ is the basis 
deformation, which varied in the interval $[-0.2,0.2]$.


\begin{table}[!htb]
\caption{\label{tab:timewise-split-rms} Root mean square error 
for the three different EDFs considered in this work. The first line of each EDF 
is for the direct DFT calculation. The following three lines are for the 
modified values after subtracting the bias trained on different sets of 
features (DA, APA, and SPA). The data is split into training, testing and 
prediction following the time-wise split as shown in Panel (a) of Fig. 2. of 
the main manuscript. All quantities are in units of MeV.}
 \begin{ruledtabular}
 \begin{tabular}{l d d d}
 Model        & \multicolumn{1}{c}{Training}
              & \multicolumn{1}{c}{Testing}
              & \multicolumn{1}{c}{Prediction} \\
 \hline
 UNEDF0                   & 1.62 & 1.68 & 1.80 \\
 UNEDF0 + DA              & 0.27 & 0.99 & 0.82 \\
 UNEDF0 + APA             & 0.26 & 0.75 & 0.77 \\
 UNEDF0 + SPA             & 0.28 & 0.91 & 0.83 \\
 SLY4                     & 3.95 & 4.22 & 5.15 \\
 SLY4 + DA                & 0.28 & 1.13 & 0.88 \\
 SLY4 + APA               & 0.28 & 0.95 & 0.80 \\
 SLY4 + SPA               & 0.31 & 1.07 & 0.88 \\
 UNEDF1$_{\rm HFB}$       & 2.85 & 3.45 & 3.10 \\
 UNEDF1$_{\rm HFB}$ + DA  & 0.22 & 0.90 & 0.77 \\
 UNEDF1$_{\rm HFB}$ + APA & 0.26 & 0.92 & 0.79 \\
 UNEDF1$_{\rm HFB}$ + SPA & 0.29 & 0.93 & 1.12
\end{tabular}
\end{ruledtabular}
\end{table}

\section{Root mean square errors of DFT predictions}

We list in tables \ref{tab:timewise-split-rms}-\ref{tab:large-z-split-rms} the 
root mean square deviation 
$
\sigma = 
\sqrt{\frac{1}{n} \sum_{i=1}^n \left[B_{\rm theo}(Z_i, N_i) - B_{\rm exp}(Z_i, N_i) \right]^2}
$
for each functional (SLy4, UNEDF0, UNEDF1$_{\rm HFB}$). For each functional, 
the tables list the rmse without any bias correction and with bias correction: 
direct-approach (DA), all-properties approach (APA) or selected-properties 
approach (SPA). The rmse are presented in the training, testing and prediction 
regions separately.

\begin{table}[!htb]
\caption{\label{tab:large-n-split-rms} Same as Table~\ref
 {tab:timewise-split-rms} but for the large-N split between training, testing
 and prediction sets as described in the text and shown in Panel (b) of Fig. 2.}
%
 \begin{ruledtabular}
 \begin{tabular}{l d d d}
 Model        & \multicolumn{1}{c}{Training}
              & \multicolumn{1}{c}{Testing}
              & \multicolumn{1}{c}{Prediction} \\
 \hline
 UNEDF0                   & 1.34 & 1.52 & 1.88 \\
 UNEDF0 + DA              & 0.25 & 0.69 & 1.41 \\
 UNEDF0 + APA             & 0.24 & 0.51 & 1.27 \\
 UNEDF0 + SPA             & 0.29 & 0.67 & 1.20 \\
 SLY4                     & 2.86 & 2.79 & 5.21 \\
 SLY4 + DA                & 0.26 & 0.83 & 1.90 \\
 SLY4 + APA               & 0.28 & 0.77 & 2.13 \\
 SLY4 + SPA               & 0.29 & 0.77 & 2.11 \\
 UNEDF1$_{\rm HFB}$       & 2.67 & 3.09 & 3.26 \\
 UNEDF1$_{\rm HFB}$ + DA  & 0.22 & 0.70 & 1.49 \\
 UNEDF1$_{\rm HFB}$ + APA & 0.25 & 0.71 & 1.57 \\
 UNEDF1$_{\rm HFB}$ + SPA & 0.27 & 0.78 & 2.14
\end{tabular}
\end{ruledtabular}
\end{table}

\begin{table}[!htb]
\caption{\label{tab:large-z-split-rms} Same as Table~\ref
 {tab:timewise-split-rms} but for the large-Z split between training, testing
 and prediction sets as described in the text and shown in Panel (c) of Fig. 2. }
%
 \begin{ruledtabular}
 \begin{tabular}{l d d d}
 Model        & \multicolumn{1}{c}{Training}
              & \multicolumn{1}{c}{Testing}
              & \multicolumn{1}{c}{Prediction} \\
 \hline
 UNEDF0                   & 1.73 & 1.78 & 1.27 \\
 UNEDF0 + DA              & 0.32 & 0.92 & 2.43 \\
 UNEDF0 + APA             & 0.26 & 0.68 & 1.64 \\
 UNEDF0 + SPA             & 0.27 & 0.73 & 1.16 \\
 SLY4                     & 2.92 & 2.61 & 6.37 \\
 SLY4 + DA                & 0.33 & 1.06 & 3.15 \\
 SLY4 + APA               & 0.30 & 0.85 & 3.03 \\
 SLY4 + SPA               & 0.40 & 0.98 & 3.40 \\
 UNEDF1$_{\rm HFB}$       & 3.17 & 3.27 & 2.43 \\
 UNEDF1$_{\rm HFB}$ + DA  & 0.25 & 0.89 & 3.51 \\
 UNEDF1$_{\rm HFB}$ + APA & 0.26 & 0.99 & 1.24 \\
 UNEDF1$_{\rm HFB}$ + SPA & 0.52 & 1.61 & 1.68
\end{tabular}
\end{ruledtabular}
\end{table}


\begin{figure*}[!ht]
\includegraphics[width=\linewidth]{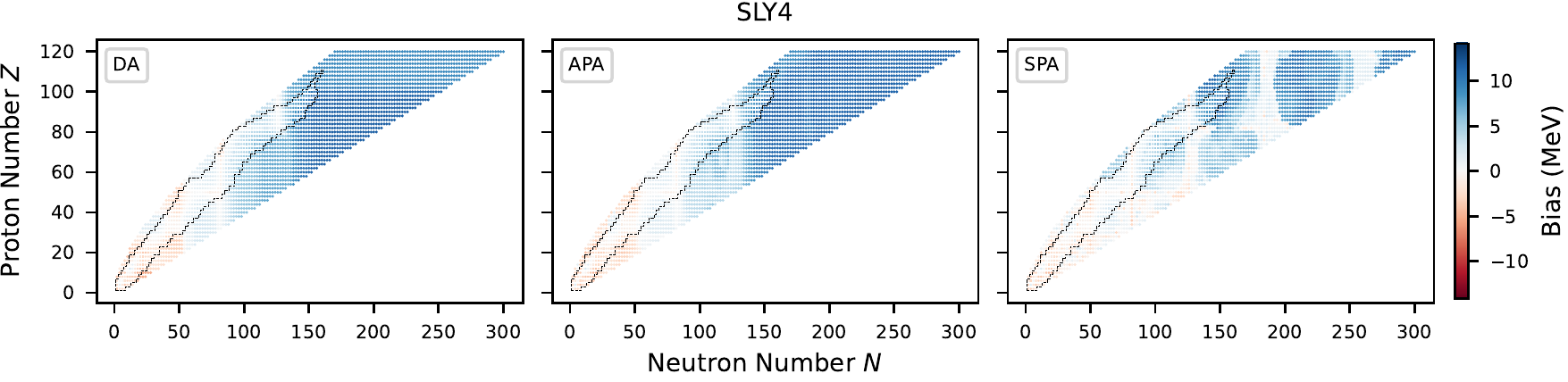}
\includegraphics[width=\linewidth]{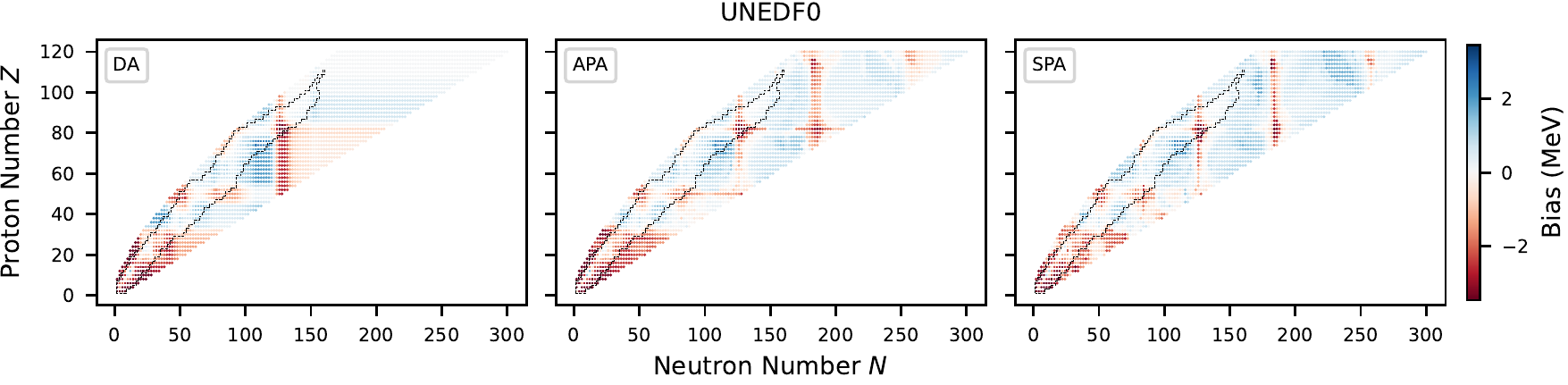}
\includegraphics[width=\linewidth]{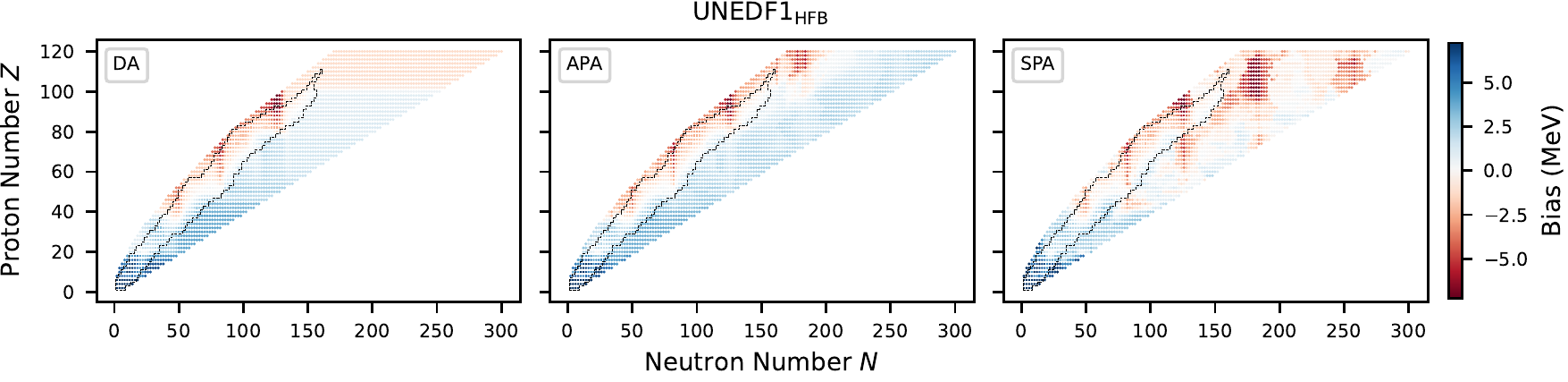}
\caption{\label{fig:bias} Value of the estimated AI/ML bias across the entire 
mass table for the SLy4, UNEDF0 and UNEDF1$_{\rm HFB}$ parameterizations of the 
Skyrme functional and the three feature selection modes.}
\end{figure*}

\section{Estimated bias}

Figure \ref{fig:bias} shows the estimated bias for the SLy4, UNEDF0 and 
UNEDF1$_{\rm HFB}$ functionals across the entire mass table after using
\emph{all} even-even nuclei in the AME2020 data set for training and testing. In the direct 
approach (DA, left column), the bias shows very little structure as soon as we 
depart significantly from the training/testing region (marked by the contour in each plot. 
In the case of SLy4, where the actual discrepancy is very large and grows approximately like 
$Z^2$, this lack of structure does not have major consequences. In the case of 
the UNEDF0 functional, which reproduces binding energies reasonably well, it 
implies that the AI/ML algorithm has not learned any dependency of the bias as a 
function of $Z$ and $N$ outside of the training/testing region. The APA approach improves things significantly, since 
structures associated with magic numbers become more visible -- although for 
UNEDF1$_{\rm HFB}$ the differences between DA and APA are still small. In the 
selected approach (SPA), the bias has a clear, non-trivial dependency on $Z$ 
and $N$ with neutron shell closures, in particular, clearly visible: the AI/ML
algorithm has learned physics trend from the available data.

\begin{figure*}[!ht]
\includegraphics[width=\linewidth]{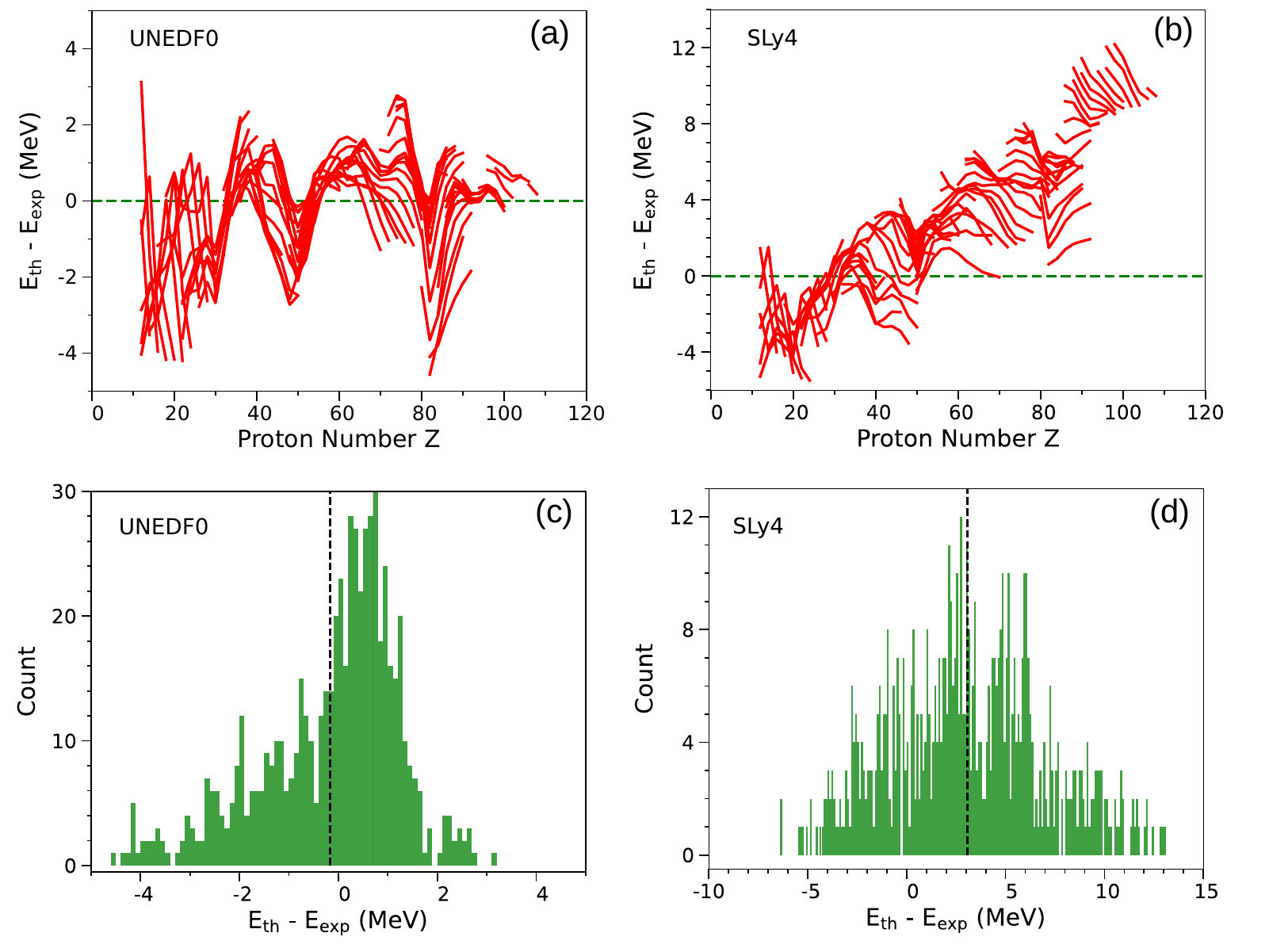}
\caption{\label{fig:masses} Residuals of nuclear binding energies as for 
isotonic sequences, as a function of proton number. Panel (a): results for 
UNEDF0; panel (b): results for SLy4. Panels (c) and (d) show the corresponding 
distributions of the residuals.}
\end{figure*}

Figure \ref{fig:masses} provides additional insights on the differences in 
the improvement index $\eta$ in the prediction region. The upper row of the 
figure shows the residuals $E_{\rm th} - E_{\rm exp}$ for isotonic chains as a 
function of proton numbers for UNEDF0, panel (a), and SLy4, panel (b). While 
both sets of results exhibit the ``traditional'' large biases around closed 
shells, the average error for the UNEDF0 is approximately centered around 0, as 
can be seen in panel (c). In contrast, the error for SLy4 is not only much 
larger, the overall error is clearly correlated with the value of the proton 
number Z. Panels (c) and (d) show the distributions of the residuals: for the 
UNEDF0 functional, this distribution is approximately centered around 0, while 
for SLy4, it is centered at around 3 MeV. The dependency of the residual on 
proton number is simple and robust: it is in fact a consequence of the fact 
that in the fit of the SLy4 functional, nuclear masses were not corrected 
for the electronic binding energy, which grows approximately like $Z^2$. As a 
result, the total discrepancy for SLy4 can be written 
\begin{equation}
\Delta E = \Delta E_{\rm elec.} + \Delta E_{\rm nucl.}.
\end{equation}
It is the sum of a discrepancy originating from the electronic energy 
correction, $\Delta E_{\rm elec.}$, and an additional term accounting for every 
other missing nuclear physics, $\Delta E_{\rm nucl.}$. In heavier nuclei, 
$\Delta E_{\rm elec.} \gg \Delta E_{\rm nucl.}$. The AI/ML model can easily 
learn the trend of $\Delta E_{\rm elec.}$ from every data set we considered in 
this work. Most importantly, this trend remains valid at large $Z$ values, 
which means that the gain from removing the bias will remain significant in the 
prediction region, contrary to UNEDF0.

\begin{acknowledgments}
Discussions with M. Grosskopf, E. Lawrence, S. Wild and J. O'Neal are gratefully 
acknowledged.
Support for this work was partly provided through Scientific Discovery through
Advanced Computing (SciDAC) program funded by U.S. Department of Energy,
Office of Science, Advanced Scientific Computing Research and Nuclear
Physics. It was partly performed under the auspices of the US Department of
Energy by the Lawrence Livermore National Laboratory under Contract
DE-AC52-07NA27344. Computing support for this work came from the Lawrence
Livermore National Laboratory (LLNL) Institutional Computing Grand Challenge
program.
\end{acknowledgments}

\bibliography{references}